# The Role of Planetary-Scale Waves on the Stratospheric Superrotation in Titan's Atmosphere

Yuan Lian*[1], Cecilia Leung[2], Claire Newman[1], Leslie Tamppari[2],

1. Aeolis Research, Chandler, AZ
2. NASA JPL, Pasadena, CA
* lian@aeolisresearch.com

**Abstract**

We analyze simulation results from the TitanWRF global circulation model to understand the mechanisms that maintain the equatorial superrotation in Titan's stratosphere. We find that the eddies associated with wave activities can transport angular momentum upgradient to zonal flow, leading to acceleration of the equatorial superrotation. The dominant wave modes identified in this study are consistent with previous studies, with zonal wavenumber 1 being the major contributor to the prograde acceleration. Despite the same conclusion of maintenance of equatorial superrotation via wave-mean interactions, we find that the way waves interact with the zonal flow in TitanWRF is slightly different from some other studies. We confirm our previous findings that in TitanWRF this occurs primarily during a dozen or so annual, short-duration (a few Titan sols) angular momentum "transfer events," which have a repeatable seasonal pattern but differ slightly in timing and magnitude between years. This is not the case in the Titan Atmosphere Model (TAM), which found milder angular momentum transfers that produced the strongest acceleration of superrotation around solstice in the upper stratosphere and more continuous year-around acceleration in the lower stratosphere. Despite differences in angular momentum transfer across models, we further find that, similar to the TAM wave analysis results, eddies generated by Rossby-Kelvin instabilities may be the major source of prograde angular momentum for the equatorial superrotation, although TitanWRF may also include contributions from the absorption of vertically propagating equatorial Kelvin waves. This differs from our previous work, which suggested barotropic waves were responsible for TitanWRF's solsticial transfer event.

**1. Introduction**

Titan, the largest moon of Saturn, has an atmosphere that resembles many aspects of the Earth, such as having a significant surface pressure (~44% larger than on Earth), being rich in nitrogen, and having both an active hydrological cycle and seasonally varying Hadley circulations. There are also, however, distinct differences in composition and in thermal and wind structures, among which the equatorial stratospheric superrotation is of particular interest. Here the superrotation refers to the atmospheric condition where the ratio between the total angular momentum of the atmosphere in motion and that of the atmosphere at rest is greater than unity. Like Venus, Titan is a slowly rotating planet, with self-rotation periods about 243 (Venus) and 16 (Titan) times longer than that of Earth. Despite this, both Venus and Titan exhibit strongly superrotating stratospheres. A well-known issue in simulating the atmospheric circulation of both bodies with 3D numerical global circulation models (GCMs) is the difficulty in reproducing the full extent of this stratospheric superrotation. For Titan's atmosphere, while the model of Hourdin et al. [1993] was able to produce moderately strong winds, that model is no longer in use, and several other models (*e.g.,* Tokano [1999], Friedson et al. [2009] and Larson et al. [2014]) have to date been unable to produce strong superrotation without artificially enhancing zonal winds. However, the models of Newman et al. [2011], Lebonnois et al. [2012], and Lora et al. [2015] all produce realistically strong winds.

The difficulty arises mainly from both the physical forcing of Titan's circulation and the fact that capturing the process that drive superrotation present an unusual challenge to the models' numerics (dynamical solvers), which were typically designed for use on Earth. Physically, Titan receives very little solar radiation but it is still able to drive the temporal and seasonally varying global dynamics such as wind, temperature and methane cycle (*e.g.*, Roe et al. [2005]; Turtle et al. [2018]; Teanby et al. [2019]). Titan's atmosphere also has a long radiative timescale in most regions. Hence, in order to reach a steady state, the model must integrate for much longer than those for Earth and Mars. Numerically, the dynamical filters and parameterizations of viscosity that are commonly implemented in atmospheric models to suppress numerical instabilities (*e.g.*, polar filters in models where the east-west grid spacing shrinks toward the poles) may interfere with the angular momentum exchanges needed to 'spin up' the winds and produce equatorial superrotation (*e.g.*, Newman et al. [2011]). Therefore, it is crucial to study the mechanisms that drive and maintain the superrotation in order to improve the realism of superrotation in Titan numerical models, especially when introducing new boundary conditions (*e.g.*, topography; McDonald et al. [2016]) or physical process (e.g., active hydrological cycles; Schneider et al. [2012], Lora et al. [2015], Faulk et al. [2020], Newman et al. [2016], Mitchell [2008], Horvath et al. [2016]).

The maintenance of equatorial superrotation on slowly rotating planets including Titan requires upgradient transport of angular momentum that counteracts diffusive processes at the equator (*e.g.*, Wang and Mitchell [2014]; Dias Pinto and Mitchell [2016]; Read and Lebonnois [2018] and references therein). Due to the weak Coriolis effect on Titan, the thermally-direct, angular-momentum conserving meridional Hadley cell circulations could extend from equator to pole during equinoxes and pole-to-pole during solstices. Hide's theorem states that there cannot be a local maximum or minimum of angular momentum in the interior of the steady-state fluid in the presence of any form of diffusion that eradicates the extrema [Hide, 1969]. This would prohibit the presence of equatorial superrotation if angular momentum were solely transported by zonally-symmetric meridional circulations, because the equatorial superrotation is equivalent to a local maximum of angular moment. Gierasch [1975] suggested that the upgradient transport of angular momentum via eddies/waves could be responsible for the maintenance of the extrema. There are two major mechanisms that may be responsible for the eddy/wave generation: eddies/waves are generated by barotropic instabilities associated with axisymmetric flow at mid-to-low latitudes (*e.g.*, Rossow and Williams [1979]); or, eddies/waves are generated by non-axisymmetric forcing such as baroclinic instabilities associated with differential heating (*e.g.*, Scott and Polvani [2008]). Planetary waves are thought to be responsible for equatorial acceleration on slow-rotating planets (see Read and Lebonnois [2018] for a comprehensive review). Depending on the propagation direction, these waves could lead to acceleration and deceleration at their sources and sinks (*e.g.*, Edmond et al. [1980]; Vallis [2006]). For instance, equatorial Rossby waves radiating away from their sources may accelerate the equatorial flow while decelerate the flow at higher latitudes where they are dissipated (*e.g.*, Suarez and Duffy [1992]; Schneider and Liu [2009]; Liu and Schneider [2011]). Another jet driving mechanism, the Rossby-Kelvin instability, is gaining popularity (Imamura [2006]; Mitchell and Vallis [2010]; Wang and Mitchell [2014]; Zurita-Gotor and Held [2018]; Takagi et al. [2022]; Lewis et al. [2023]). This mechanism depicts the scenario where equatorial acceleration is a result of the resonant coupling between equatorially trapped Kelvin waves and mid-latitude Rossby waves, *i.e.*, the instability induced by resonance between the waves can break and dissipate Kelvin waves thus producing prograde acceleration at the equator.

Previous studies performed theoretical and numerical wave analysis in order to understand the conditions under which eddies could lead to acceleration of the equatorial wind in Titan's stratosphere, since eddies can be considered as superpositions of waves at various wavelengths and frequencies. Newman et al. [2011] analyzed Titan Weather Research and Forecasting (TitanWRF) GCM results and found that equatorial wind speeds increased mainly during short (a few Titan sols) "angular momentum transfer events" which occurred with a repeatable seasonal pattern, although the exact timing and magnitude of the events differed each year (even after the model was fully 'spun up'). During these events, they identified critical layers where the phase speeds of dominant wave modes intersect with zonal winds at low-to-mid latitudes. They interpreted this as the convergence and divergence of the eddy momentum fluxes, associated with the eddies generated by breaking of meridionally propagating waves near critical layers, leading to the acceleration and deceleration of zonal winds at the equatorward and poleward side of the critical layers, respectively. Lewis et al. [2023] analyzed results from the Titan Atmosphere Model (TAM) GCM [Lora et al., 2015; Lombardo and Lora, 2023a, 2023b] and found that, using the theoretical considerations of equatorial Rossby-Kelvin waves [Wang and Mitchell, 2014], the equatorial superrotation was driven by the eddies generated by the Rossby-Kelvin instabilities under resonance conditions between the two types of the waves. They also did not identify strong, short-lived angular momentum transfer events in TAM. Instead, they found mild angular momentum transfers that produced the strongest acceleration of superrotation around solstice in the upper stratosphere and more year-around acceleration in the lower stratosphere. Due to the discrepancy between the two models in terms of the jet driving mechanisms, we perform re-analysis of Newman et al. [2011] simulation results with more complete diagnostics tools to better understand the wave-mean interactions in TitanWRF and how they compare to the TAM results. We note that a second Titan GCM we have developed, based on the MITgcm [Adcroft et al. 2004], exhibits angular momentum transfer events very similar to those in TitanWRF, despite utilizing an entirely different dynamical core. Additionally, preliminary TitanWRF simulation results with methane moist convection and topology also showed similar angular momentum transfer events over multiple years (see Supplemental Material for both cases). However, as the jet-driving theories relevant to Titan were mostly originated from idealized model atmosphere (such as the mechanisms based on Hide's theorem and Rossby-Kelvin instabilities), we focus on comparing the TitanWRF (dry without topography) and TAM results in this work, and leave a deeper exploration of all aspects of the Titan MITgcm and TitanWRF with more complete physics to future work.

The paper is laid out as follows. In section 2, we provide a brief review of TitanWRF including the model configuration used here. In section 3, we present the wave analysis methods, potential atmospheric instabilities, and wave generation mechanisms. In section 4, we present wave analysis results and briefly compare our results to those in Lewis et al. [2023]. In section 5, we conclude.

## 2. Model description

### 2.1 The TitanWRF Model

The TitanWRF model in this study is very similar to that used by Newman et al. [2011]. As detailed in that paper, the model solves the system of primitive equations using the finite difference method for a dry atmosphere. It contains several physics parameterizations of physics processes that are suitable for Titan's atmosphere, such as two-stream radiative transfer with seasonal variation of solar insolation, surface and subsurface schemes that assume constant soil properties, and a planetary boundary layer that vertically mixes heat and momentum in the unstable atmosphere that is typically defined by the bulk Richardson number $R_i < 0.25$. Here $R_i$ is a dimensionless number that describes the ratio between thermal buoyancy and vertical shear of horizontal wind. The only difference in this work is the use of an updated version of WRF's official release as the base for TitanWRF, as described below. Methane cycle (*e.g.*, Newman et al. [2016]) is ignored for this study but will be considered in future for a more comprehensive inter-model comparison between TitanWRF and TAM.

For this work the planet surface is assumed to be flat, as in Newman et al. [2011], although TitanWRF may be run with topography (e.g., McDonald et al. [2016]). The vertical domain consists of 55 layers spanning $1.44 \times 10^5$ to 0.05 Pa (equivalent to the altitude from the ground up to about 500 km). The model has a horizontal resolution of 64x36, giving a grid resolution of 5.625° in longitude and 5° in latitude. This horizontal resolution is sufficient to cover the wavelengths of interest in this study. Newman et al. [2011] suggested that, in order to obtain superrotation with zonal wind profiles comparable to the Huygens and CIRS measurements (*e.g.*, Achterberg et al. [2011]; Folkner et al. [2006]), the top of the TitanWRF model may need to be extended to a higher altitude, e.g., 600 km, so that the damping layers near the model top (designed to prevent spurious wave reflections) do not remove excessive amount of angular momentum. However, extending the model top to 600 km would require modifications of the radiative transfer scheme to properly represent the photochemical processes in Titan's thermosphere.

We run the model from a spun-up, near steady state using a "restart file" from 74 Titan years into the main simulation of Newman et al. [2011]. We term it "near steady state" because, although the total angular momentum budget is maintained to be nearly constant and there is a mostly repeatable seasonal cycle, the angular momentum transfer events still exhibit annual variations. This was demonstrated in Newman et al. [2011], who compared the timings and magnitudes of the transfer events in years 75 and 76 of the original simulation in their Figure 13 and Figure 14. In fact, when we reran the Titan year 75 simulation, we found that the timings of the transfer events were slightly different from those of year 75 in the original simulation (Figure 1 here). This was due to a minor update of the underlying WRF version from 3.0 to 3.3, which included small bug fixes in the dynamical core, but did not change the overall nature of the Titan dynamics. This indicates that the exact timing and magnitude of the transfer events is very sensitive to small perturbations in model setup and initial conditions. However, the seasonal timing and nature of the events is consistent in all cases. For examining the timing of the angular momentum transfer events, the typical duration and output frequency of our simulations are 1 Titan year (equivalent to 29 Earth years) and 1 Titan day (equivalent to about 16 Earth days) respectively. For wave analysis (see Section 3.1), the typical duration and output frequency are 1 Titan day and 10 Titan minutes (equivalent to about 9600 Earth seconds) respectively.

## 2.2 Angular Momentum Transfer Events

Newman et al. [2011] suggested that the formation and maintenance of equatorial superrotation in Titan's stratosphere were closely related to the eddy momentum fluxes (hereafter, EMF) associated with barotropic instabilities. They used global and seasonal distributions of angular momentum to demonstrate how the zonal wind was accelerated at low latitudes and decelerated at mid-to-high latitudes during the sporadic occurrences of angular momentum transfer events (*i.e.*, the spikes of rate change of angular momentum in the stratosphere in Fig. 1). Here, the angular momentum in each grid cell *i* with respect to the planet self-rotation axis is defined in the same way as in Newman et al. [2011]

$$M_i = \delta m_i(u_i + \Omega a cos\phi_i) a cos\phi_i$$

where $\delta m_i$ is the mass, $u_i$ is the zonal wind, and $\phi_i$ is the latitude of the grid cell *i*. $\Omega$ is the planet rotation rate and $a$ is the planet radius. The strength of superrotation in the atmospheric layer of interest is characterized by a local superrotation index (*e.g.*, Read and Lebonnois [2018])

$$s = \frac{\sum_i^N \delta m_i(u_i + \Omega a cos\phi_i) a cos\phi_i}{\sum_i^N \delta m_i \Omega (a cos\phi)^2}$$

where $N$ is the total number of grid cells in the atmospheric layer bounded by pressure ranges (see top panel in Fig. 1). The atmosphere is superrotating if $s > 1$. Newman et al. [2011] further utilized cross-spectra analysis of EMF for a typical transfer event near $L_s = 270°$ and a "gap period" (a period separated from any transfer events) near $L_s = 278°$ to explore the mechanisms of wave-mean interactions that were responsible for acceleration and deceleration of zonal wind. However, their analysis only focused on a single transfer event and did not examine how the transfer events correlated with other physics quantities such as temperature or geopotential.

Here, we analyze a series of major transfer events restarted from year 74 of Newman et al.'s spun-up simulation, which we refer to as our year 75 (and which differs slightly from the original year 75, as described above). Despite the slight differences in the simulation (*e.g.*, exact timing and strength of the transfer events) and in the analysis performed here (*e.g.*, wave analysis tools and methods), the principles of how angular momentum transfer events drive the stratospheric superrotation in TitanWRF remain the same after detailed wave-mean analysis.

In this paper, we will focus on two major angular momentum transfer events (shown by the magnitude of the spikes in Fig. 1) in the region above 110 mb (*i.e.*, stratosphere or namely the upper atmosphere in Newman et al. [2011]). One event is at $L_s \sim 261°$ and is comparable to the solsticial event near $L_s = 270°$ in Newman et al. [2011], and we designate as a 'solsticial' event. The other is near $L_s \sim 191°$, which we designate as an 'equinoctial' event, a type of event that was not examined by Newman et al. [2011].

## 3. Wave analysis and atmospheric instabilities

### 3.1 Wave Analysis Methods

Understanding how waves and the mean flow interact requires tools to find the properties of the waves, including their spatial scales, frequencies, and how they propagate in Titan's atmosphere. In particular, we are interested in the dominant wave modes that either accelerate or decelerate the stratospheric superrotation. Here, we use several wave analysis techniques including cospectra (as in Newman et al. [2011]), a 1-dimensional (1D) Fast-Fourier Transformation (FFT), spherical harmonics (for displaying eddy fields associated with waves and decomposition of horizontal flow to divergent and rotational components), as well as some linearized wave analysis methods.

The cospectra method (also known as cross-spectra method, *e.g.*, Randel and Held, [1991]) describes the acceleration and deceleration of zonal wind via the convergence and divergence of EMF associated with waves at various wavelengths or phase speeds. Conditions for wave absorption may be assessed via this analysis method, *i.e.*, the locations where wave phase speed is close to the zonal wind speed (see section 3.2.4 for description of wave propagation near critical layers). This wave analysis method also provides the latitude-altitude distribution of dominant wave modes characterized by the magnitude of power spectral density (PSD) calculated either from EMF or from eddy heat fluxes. The cospectral analysis is performed using 1 Titan day with output interval of 10 Titan minutes. The duration is long enough to cover at least several wave periods for the dominant wave modes (*e.g.*, Newman et al. [2011]; Lewis et al., [2023]), and the output frequency is sufficient to resolve sound waves without aliasing. The details of the method are described in Appendix B.

The 1D FFT is used primarily to reveal the vertical structure of perturbation dynamical variables (such as temperature) associated with the angular momentum transfer events. This analysis shows the dominant vertical wave modes and may be used with a simple hydrostatic wave dispersion relation to establish some connection with the horizontal wave modes during the same time period, which can provide additional information on the waves of interest (*e. g.*, Wheeler and Nguyen [2015]).

Spherical harmonics can be used to decompose the dynamical field to various spatial scales in terms of wavenumbers, including the zonal-mean component whose zonal wavenumber is zero. Similar to the cospectra method, spherical harmonics analysis has been widely used to diagnose wave-wave and wave-mean interactions on Earth and gas giant planets, both in observational data (*e.g.*, Boer and Shepherd,[1983]; Galperin et al., [2014]; Read et al., [2018]; Read et al., [2020];

Siegelman et al., [2022]) and in output from numerical simulations (e.g., Schneider and Liu [2009]; Burgess et al., [2013]). Here we will use the spherical harmonics to assist the cospectra analysis by displaying geopotential anomaly at the zonal wavenumber of interest, as well as the eddy wind fields corresponding to this zonal wavenumber. Additionally, we use Windspharm, a Python spherical harmonics analysis package that analyzes the global wind field in spherical geometry (Dawson [2016]), to decompose the horizontal wind field to its divergent and rotational components, using which we can further examine the wave-mean interactions (*e.g.*, Zurita-Gotor et al., [2019, 2022], see Appendix C).

**3.2 Atmospheric instabilities**

The excitation of atmospheric waves and generation of eddy activities require perturbations to atmospheric motions. Newman et al. [2011] demonstrated that the waves likely had barotropic origin but did not analyze other wave generation mechanisms such as inertial and baroclinic instabilities, all of which are likely present on slowly rotating planets (*e.g.*, Iga et al. [2005]).

**3.2.1 Barotropic and baroclinic instabilities in QG flow**

Atmospheric instabilities are commonly studied in quasi-geostrophic (QG) flows that simplify the wave equations [Holton, 2004]. The QG approximation assumes that the Coriolis force and pressure gradient are almost in balance, implying small Rossby number that is defined as $R_o = \frac{U}{fL}$, where $U$ is a typical wind speed, $f = 2\Omega sin\phi$ is the Coriolis parameter, $L$ is a typical dynamic length scale (*e.g.*, zonal wavelengths of planetary waves). However, this approximation may not be applicable to most portions of Titan's stratosphere [Mitchell and Lora, 2016]. This is because, while Titan is a slow rotator (hence has a weak Coriolis effect), the stratosphere is superrotating with wind speeds of more than $150\ ms^{-1}$. The implication of this is that the stratosphere has a very large Rossby number, roughly on order of 10 at low latitudes even with zonal wavenumber 1 (*i.e.*, $L = 2\pi a$, where $a$ is Titan's radius), and for such large values the QG approximation breaks down, as the winds are most likely in cyclostrophic balance where the centrifugal force balances the pressure gradient. At mid-to-high latitudes, however, the Rossby number becomes less than unity due to slower wind speeds and a larger Coriolis effect. The QG approximation may apply in this case. In the lower troposphere, the QG approximation may even apply to low latitudes since the typical wind speed is merely a few meters per second, which implies a very small Rossby number compared to that in the stratosphere.

The barotropic instability criterion in Newman et al. [2011]'s analysis is 2D in nature. It depicts how the atmosphere becomes unstable when the curvature of the zonal wind $\frac{\partial^2 u}{\partial y^2}$ exceeds the meridional gradient of the Coriolis parameter β in a 2D barotropic flow (*e.g.*, on isobaric surfaces where the lateral temperature gradient is very small). The equation describing the 2D barotropic instability criterion is $\frac{\partial^2 u}{\partial y^2} - \beta > 0$, where $u$ is the zonal wind, $y$ is the distance in latitude and $\beta = \frac{\partial f}{\partial y}$ is the meridional gradient of the Coriolis parameter.

A more generalized barotropic stability criterion, the Charney-Stern theorem, defines how flow may become unstable in a 3D QG regime where stratification of the atmosphere is considered. The necessary condition for instability is that the meridional gradient of potential vorticity (PV) changes sign with latitude (Read et al. [2006]; Dowling et al. [2019]) on isentropic surfaces. The PV on isentropic surfaces (IPV, or Ertel's PV) is defined as

$$q = -g(f + \zeta_\theta)\frac{\partial \theta}{\partial p}$$

where g is the gravity, $\zeta_\theta$ is the relative vorticity on isentropic surface, $\theta$ is the potential temperature and $p$ is the pressure. The IPV may be calculated at altitudes above the planetary boundary layer (PBL), but not within the PBL where the atmosphere tends to be neutrally stratified, meaning that isentropic surfaces are ill-defined. To circumvent this issue, we use quasi-geostrophic potential vorticity (QGPV) on isobaric surfaces, which is defined as (following Read et al., [2006])

$$q_G = f + \zeta - f\left\{\frac{\partial}{\partial p}\left(\frac{p\Delta T(x,y,p)}{s(p)\overline{T}(p)}\right)\right\}$$

where $\overline{T}(p)$ is the horizontal mean temperature, $\Delta T(x,y,p) = T(x,y,p) - \overline{T}(p)$, and $s(p)$ is the stability parameter defined as $s = -\frac{\partial ln(\overline{\theta})}{\partial lnp}$, where $\overline{\theta}$ is the horizontal mean of potential temperature.

The meridional gradient of QGPV may also be used to evaluate conditions for baroclinic instability, which can be considered as a shear instability induced by lateral temperature contrasts. The conditions for baroclinic instability may be defined by four Charney-Stern-Pedlosky (CSP) necessary conditions (Polichtchouk and Cho [2012]; Read et al., [2020]), among which a pair of conditions are more relevant to our simulated Titan's atmosphere. The first one is that the IPV/QGPV gradient changes sign in the interior of the fluid flow (*e.g.*, $\partial_y q_G$ changes sign between two adjacent vertical layers), which is usually met for Earth's atmosphere due to strong beta effect and large spatial temperature contrast. For Titan, the planetary rotation rate is very small, hence QGPV is more influenced by relative vorticity (*i.e.*, $\zeta$ dominates $q_G$). The second necessary condition states that the meridional QGPV gradient has the same sign as the vertical zonal wind shear at the lower boundary. We will examine whether either or both of these baroclinic instability criteria are met in Titan's stratosphere. The rest of the CSP necessary conditions are related to the model top, but in TitanWRF the upper layers of the model are under the influence of a so-called "sponge layer" in which zonal wind/temperature is relaxed back to zonal mean value in order to prevent spurious wave reflections. Therefore, we do not examine those conditions.

### 3.2.2 Ageostrophic barotropic instability (Rossby-Kelvin Instability)

Recent studies on the generation and maintenance of superrotation on slow-rotating planets showed that the resonance modes of coupled Rossby and Kelvin waves could excite atmospheric instabilities that contribute eddy momentum to zonal wind in a barotropic flow, i.e., the Rossby-Kelvin instability (hereafter R-K instability, Sakai [1989]) or ageostrophic barotropic instability (e.g., Iga [2005]; Wang and Mitchell [2014]). More specifically, the instability instigated by Rossby-Kelvin mode 1 in Wang and Mitchell [2014]'s study was found to be the major source of eddy momentum that drove equatorial superrotation. Lewis et al. [2023] examined such instability in their TAM GCM simulation results and found that, similar to the linear wave analysis by Wang and Mitchell [2014], the convergence of eddy momentum associated with Rossby-Kelvin waves could drive stratospheric superrotation, without the wave absorption mechanism suggested by Newman et al. [2011]. Here, we perform a similar analysis to examine whether the R-K instability also presents in TitanWRF simulation results and impacts the superrotation.

There are two prerequisites for R-K instability to occur (e.g., Wang and Mitchell [2014]; Zurita-Gotor and Held [2018]; Lewis et al. [2023]). First, some spatial overlap between Rossby and Kelvin waves is required. This requirement is easily satisfied, as the equatorial Rossby deformation radius is comparable to the planet radius in Titan's stratosphere for typical wind speed $U > 150 ms^{-1}$, enough to create spatial overlap between equatorial Kelvin waves and mid-latitude Rossby waves (see Lewis et al. [2023] for a detailed consideration on the equatorial Rossby deformation radius and meridional extent of the waves). The second requirement is that both types of the waves have similar frequencies so they can resonate. We use the resonant condition established by Iga and Matsuda [2005] and define a Froude number similar to Wang and Mitchell [2014]

$$F_r = \frac{\widehat{\omega}_R + k_x^R \overline{U}_R}{\widehat{\omega}_K + k_x^K \overline{U}_K} \approx \frac{\overline{U}_{mid}/cos\phi_{mid}}{c_0^K}$$

This Froude number, the ratio between the Doppler-shifted mid-latitude Rossby wave frequency and the Doppler-shifted equatorial Kelvin wave frequency, is defined to evaluate the presence of such a condition in Titan's atmosphere. In the above equation, the subscript $R$ and $K$ denote Rossby and Kelvin waves respectively, $\widehat{\omega} = \omega - k_x\overline{U}$ is the intrinsic wave frequency and $\omega$ is the wave frequency. Wang and Mitchell [2014] as well as Lewis et al. [2023] assumed that the intrinsic component of the Rossby wave frequency $\widehat{\omega}_R$ was small and could be ignored. This assumption is consistent with the typical properties of Rossby waves (see equatorial Rossby wave properties as an example in section 4.3). We adopt a similar approach and choose the wind in the vicinity of the critical layers on horizontal planes. Critical layers exist at latitudes (typically near the mid latitudes) where the wave phase speed becomes comparable to the zonal wind speed, meaning that the intrinsic Rossby wave phase speed (i.e., the wave's speed relative to the background flow) is close to zero. Hence, we select mid-to-high latitudes where the intrinsic Rossby wave phase speed is small (about an order of magnitude smaller than the background zonal flow). $\overline{U}_{mid}$ is the absolute value of the maximum phase speed for Rossby waves, $\phi_{mid}$ ( 45° N for $L_s \sim 261°$ and 45° S for $L_s \sim 191°$) is the latitude corresponding to $\overline{U}_{mid}$, and $\cos\phi_{mid}$ is to factor in the variation of zonal wavelength as a function of latitude. $c_0^K$ is the phase speed of the equatorial Kelvin wave relative to the planetary frame ($190 ms^{-1}$ for $L_s \sim 261°$ and $130 ms^{-1}$ for $L_s \sim 191°$). The zonal wave modes $n_x$ of Rossby and Kelvin waves must be the same for them to interact (Sakai [1989]). Here $n_x$ is the number of zonal waves around a latitude circle (not to be confused with zonal wavenumber that is $k_x = \frac{n_x}{a\cos\phi}$, where $a$ is the planet radius). This formula only provides a qualitative measurement of R-K instability since we would need a wave model similar to that in Wang and Mitchell [2014] to evaluate the growth mode rigorously.

### 3.2.3 Inertial instability

The inertial instability stems from the imbalance between pressure gradient and centrifugal forces. For Titan's stratosphere, such imbalance likely occurs at low latitudes where the Coriolis effect is negligible. The criterion for inertial instability can be described as that the system is unstable if the angular momentum increases towards the poles when the background flow is in cyclostrophic balance. The inertial instability occurs when $f(\zeta + f) < 0$ for an axisymmetric vortex (*i.e.*, a vortex with rotational symmetry around its rotation axis). This condition means that a parcel in a vortex would accelerate radially away from its original position when the pressure gradient and inertial forces are out of balance (Holton [2004]; Eliassen and Kleinschmidt [1957]). For northern hemisphere, this means that

$$-\frac{1}{a\cos\phi}\frac{\partial M}{\partial \phi} = \zeta + f < 0$$

where $a$ is the planet radius, $M = a\cos\phi(u + \Omega a \cos\phi)$ is the absolute angular momentum, and $u$ is the zonal wind.

**3.2.4 Wave propagation at critical layers**

Planetary waves propagating away from their sources may encounter regions where the background atmospheric flow speed becomes comparable to the wave phase speed. These regions are called critical layers when the wave phase speed and the background flow speed become equal. Waves at critical layers can break and be absorbed, dissipated, or reflected, depending on the atmospheric conditions (Killworth and McIntyre [1985]). Using the vertically propagating internal gravity waves in an isothermal, Boussinesq flow with constant wind speed as an example, the dispersion relation for these kinds of waves can be expressed as (*e.g.*, Nappo [2002])

$$k_z^2 = \frac{N^2}{(U-c)^2}$$

where $N$ is the Brunt Väisälä frequency, $k_z$ is the vertical wavenumber, $U$ is the wind speed and $c$ is the zonal wave phase speed. When $c = U$, the above dispersion relation will be a singularity because the denominator becomes zero. A similar dispersion relation may be formulated for meridionally propagating waves. The singularity near the critical line where $c = U$ will make the wave unstable because the vertical wavenumber $k_z$ becomes infinitely large (infinitesimal vertical wavelength). Waves break (are irreversibly deformed) near the critical layers, where they may be absorbed by the background flow (*i.e.*, wave energy/momentum flux deposited to the background flow), dissipated due to diffusive processes (*e.g.*, eddy viscosity), or reflected (since a critical layer also behaves as an evanescent layer) [Haynes, 2015]. Here the wave "evanescence" means a dynamically or statically unstable atmosphere in which wave amplitude decays to zero rapidly as the unstable atmosphere impairs vertical propagation of waves [Lindzen and Tung, 1976; Lindzen, 1981]. The effect of wave reflection becomes significant near a nonlinear critical layer when dissipation is not present [Haynes, 2015]. Here the nonlinear motion near a critical layer may be a result of wave breaking or other sources of instabilities such as the aforementioned R-K instability.

Two popular methods to identify the transmission, absorption and reflection of planetary waves near critical layers are the Eliassen-Palm (EP) flux diagnostics and the QGPV gradient diagnostics. The EP flux is closely related to the eddy momentum flux (see Appendix A), and its divergence and convergence can be used to infer the sources and sinks of waves (Holton [2004]). Waves are transmitted if the strength and orientation of EP flux vectors are maintained, reflected if EP flux vectors turn against their incoming direction while maintaining their strength at reflective surfaces (*e.g.*, critical layers), and absorbed near critical lines where EP flux divergence or convergence shows local maxima (Kanzawa [1982]; Brunet and Haynes [1995]; Lu et al., [2017]; Matthias and Kretschmer [2020]).

Alternatively, the transmission, absorption and reflection of waves may be determined by meridional gradient of QGPV with prior knowledge of wave properties such as horizontal wavelength and phase speed (see Appendix E). This method states that, for retrograde waves in an atmosphere where the properties of background atmosphere vary slowly compared to the vertical wavelengths of interest, the region with $\partial_y q_G \equiv \frac{\partial q_G}{\partial y} > 0$ allows wave transmission and the region with $\partial_y q_G < 0$ results in wave absorption/reflection, and vice versa for prograde waves (Andrew et al. [1987]; Charney and Drazin [1961]; Harnik and Lindzen [2001]). Mathematically, the conditions for wave transmission and absorption/reflection can be described by a refractive index $n_{ref}^2 \propto -\frac{\partial_y q_G}{c-\bar{u}}$ (Appendix E), with $n_{ref}^2 > 0$ for transmission, and $n_{ref}^2 < 0$ for absorption/reflection at a reflecting surface defined by $n_{ref}^2 = 0$ (Harnik and Lindzen [2001]).

Additional scenario near the critical layers relevant to TitanWRF is wave dissipation due to numerical difficulties. GCMs are not capable of resolving critical layers due to finite grid resolution. When waves approach the critical layers, they will instead produce grid-scale oscillations in the form of eddies or numerical noise. These eddies then deposit their momentum to the background flow. At the same time, numerical dampers designed to suppress the growth of spurious oscillations dissipate the eddy momentum. In GCMs, the behavior of wave activities at critical layers may be parameterized to represent the wave-induced force due to wave breaking at scales much smaller than the grid resolution [Lindzen, 1981].

## 4. Results

In this section, we present our re-analysis of TitanWRF results in several aspects, including both basic diagnostics of momentum forcing on the zonal flow and jet acceleration via wave-mean interactions. First, we show the contribution of mean and eddy acceleration to the zonal flow. Second, we perform cospectral analyses similar to those in Newman et al. [2011], except that the analysis now includes two major angular momentum transfer events, near autumn equinox ($L_s \sim 191°$) and northern winter solstice ($L_s \sim 261°$), with the latter being a direct comparison to the $L_s = 270°$ transfer event in Newman et al. [2011]. The zonal winds during these two transfer events are displayed in Fig 2. Each event lasts about 3-4 Titan days. The former event occurs at a pressure range between 20-2000 Pa while the latter occurs at a pressure range between 10-200 Pa. Third, we diagnose the zonal wind acceleration and deceleration due to wave-mean interactions using the information obtained via cospectral analysis and spherical harmonics decomposition of horizontal flow. FFT analysis of vertical wave modes during these transfer events is then used to examine vertical wave modes that may be associated with the dominant horizontal waves at the equatorial region. Finally, we show the diagnostic results of various atmospheric instabilities that may be responsible for the generation of the atmospheric waves.

**4.1 Jet acceleration by eddies**

The equatorial superrotation in the stratosphere is primarily driven by eddy-mean interactions, as implied by the correlation between zonal wind tendency and the divergence of EP flux that is also indicative of wave activities. A commonly used method to describe eddy-mean interactions is the transformed Eulerian mean (TEM) equations (see Appendix A). We follow the definition of the TEM terms in Newman et al. [2011] and Andrews et al. [1987]. As in Eq. 4 of Newman et al. [2011], term 1 is the total zonal wind acceleration $\frac{\partial \bar{u}}{\partial t}$, term 2 includes both the Coriolis effect and advection in terms of the meridional residual mean wind, term 3 is the vertical advection of zonal momentum due to residual mean circulation, term 4 is the divergence of EP flux, and term 5 is the dissipation of momentum due to viscosity etc. Figure 3 shows the TEM terms 1-4 in spherical coordinates during the $L_s \sim 261°$ transfer event. The zonal wind tendency $\frac{\partial \bar{u}}{\partial t}$ (Fig. 3a) shows that an eastward acceleration dominates in the equatorial region from 10 Pa to 200 Pa. This acceleration correlates well with the divergence of EP flux (Fig. 3d), suggesting that the zonal wind is mostly driven by the convergence of EMF. In the region above 10 Pa, the zonal wind experiences mostly deceleration in both hemispheres. The residual circulation is the major contributor to the acceleration in the northern hemisphere and deceleration in the southern hemisphere (Fig. 3b and 3c), while the divergence of EP flux counteracts with it (Fig. 3d). The TEM term 5 is not trivial in the cases where dissipation is strong. For instance, there is a notable imbalance among the zonal wind tendency, the acceleration due to the residual circulation, and the EP flux towards the model top, where the sponge-layer damping is strongest (not shown). Other sources of the TEM term 5 are either numerical (*e.g.*, hyperviscosity that diffusively suppresses spurious oscillations in the fluid flow and works preferentially on the grid scale) or physical (*e.g.*, diffusive PBL mixing). The region deeper than 200 Pa is mostly quiescent, suggesting very little eddy-mean interaction, implied by the near-zero divergence of EP flux ($\nabla \cdot F \approx 0$).

The effect of eddies on the zonal flow exhibits seasonal variabilities. Fig. 4 shows the TEM terms during the $L_s \sim 191°$ transfer event. Overall, the acceleration due to TEM terms is a few times weaker than that during the $L_s \sim 261°$ transfer event, as expected with a weaker angular momentum tendency during the event (see Fig. 1). The divergence of EP flux (Fig. 4d) dominates the jet acceleration at equatorial latitudes, and deceleration at mid-latitudes, between 1000 Pa and 3000 Pa, while the meridional residual circulation (Fig. 4b) dominates the rest of the stratosphere up to a few pascals.

**4.1.1 EP flux and its implication on wave activities**

The EP flux vectors ($F_y$,$F_z$) can describe the movement of wave activities [Andrews et al., 1987] from their sources (corresponding to EP flux divergence) to sinks (corresponding to EP flux convergence). Note that the orientations of EP flux vectors do not represent the actual wave propagation direction without prior knowledge of intrinsic phase speed (Andrews et al., [1987]; Imamura [2006]; Lewis et al., [2023]). Therefore, we use EP flux divergences and wave energy flux directions to evaluate wave sources and propagation, reserving the total EP flux vectors to describe other wave activities unless specified otherwise. During $L_s \sim 261°$ transfer event, the EP flux originating from the troposphere (mostly near the equator and southern low latitudes) moves upwards and gets partially deflected at about 100-200 Pa before it turns to the northern mid-latitudes, where there appears to be a potential source of waves implied by strong and mostly vertical EP flux divergence near 100 Pa and 50° N (Fig. 3d). For $L_s \sim 191°$ transfer event, the EP flux originating from the southern low latitudes in the troposphere terminates at much lower altitude near 6000 Pa, where it meets the downward EP flux originating from the equatorial region in the pressure range between 1000 Pa and 3000 Pa and gets deflected towards the northern low latitudes (Fig. 4d). Like $L_s \sim 261°$ transfer event, there is also a potential wave source at midlatitudes, which is again implied by the EP flux divergence near 1500 Pa and 55° S. The northward and downward deflection of upward EP fluxes near 100-200 Pa for $L_s \sim 261°$ transfer event implies partial wave reflection due to wave breaking in the vicinity of critical layer (*e.g.*, Brunet and Haynes [1995]; Lu et al., [2017]; Matthias and Kretschmer [2020]). For $L_s \sim 191°$ transfer event, the likelihood of wave reflection near 6000 Pa is yet to be determined since the critical layer sits at much higher altitude (~ 1000 Pa, see Figure 10a and Section 4.4). The complex EP flux patterns in the region between 1000 Pa and 6000 Pa could result from the superposition of waves from multiple sources or interference among reflected waves, which require further investigation in a future study. The vertical propagation of equatorial waves including their transmission, absorption and reflection properties will be presented again in Section 4.5 in the context of QGPV.

The orientations of the EP flux vectors at the equatorial region are dependent on the wave types, and they can visually describe wave propagation in combination of wave energy flux for either prograde or retrograde waves. The relation between EP flux and group velocity was established by previous studies (e.g., Edmond et al. [1980]; Andrews et al. [1983])

$$\vec{F} = \vec{c_g} A$$

where $\vec{F} = (F_\phi, F_z)$ is the EP flux vectors (see appendix A), $\vec{c_g} = (\hat{c}_g^\phi, \hat{c}_g^z)$ is the intrinsic group velocity, and A is the wave activity density that can be described as $A = -E/\hat{c}$ (Kawatani et al. [2010]) with E being wave energy per unit mass and $\hat{c} = c - \bar{u}$ being the intrinsic zonal phase velocity. Note that the negative sign in front of $E/\hat{c}$ arises from the different definition of EP flux components, as the convention used in Kawatani et al. [2010] has the opposite sign compared to ours (see equations 6 and 8 in Kawatani et al. [2010] and A1, A5 and A6 in Appendix A). Since E is always positive, the orientation of $\vec{F}$ is determined by the signs of $\hat{c}_g^\phi/\hat{c}$ and $\hat{c}_g^z/\hat{c}$. In the vertical direction, this means positive (upward) group velocity $\hat{c}_g^z$ and positive $\hat{c}$ (prograde wave) produce a negative (downward) flux $F_z$. Similarly, negative (downward) group velocity $\hat{c}_g^z$ and negative $\hat{c}$ (retrograde wave) also produce a negative (downward) flux $F_z$.

The vertical wave propagation is of particular interest since it provides information about which part of the atmosphere waves may originate from. Instead of directly calculating the vertical group velocity, we use wave energy flux $\overline{p'w'}$ to evaluate the sign of vertical group velocity $\hat{c}_g^z$ since wave energy propagates at group velocity. The vertical wave energy flux itself may be approximated via Eliassen and Palm's first theorem where $\overline{p'w'} = \rho_o(c - \bar{u})\overline{u'w'}$ (Eliassen and Palm [1961]; Lindzen [1990]). For prograde waves, the vertical EMF has the same sign as the vertical wave energy flux (or equivalently the vertical group velocity), *i.e.*, $sign(\overline{u'w'}) = sign(c_g^z)$ and vice versa for retrograde waves, *i.e.*, $\text{sign}(\overline{u'w'}) = -\text{sign}(c_g^z)$. That is, for both prograde and retrograde waves, positive or negative $\overline{u'w'}$ corresponds to downward or upward EP flux respectively if Eliassen and Palm's first theorem stands.

The orientations of the vertical EP flux vectors are mostly consistent with the sign of vertical wave energy flux inferred from the vertical EMF at the low-to-mid latitudes (within $\pm 60°$). Figure 5 shows the vertical EMF $\overline{u'w'}$ (multiplied by a factor of $\bar{\rho}a\cos\phi$) and EP flux vectors for zonal wavenumber 1 ($n_x = 1$) eddies with their prograde and retrograde components during both $L_s \sim 261°$ transfer event (Fig. 5a and 5b) and $L_s \sim 191°$ transfer event (Fig. 5c and 5d). During $L_s \sim 261°$ transfer event, $\overline{u'w'}$ for $n_x = 1$ prograde waves is mostly positive from ground up to 10 Pa near the equator and negative in the atmospheric layers deeper than 2000-3000 Pa at the northern low latitudes (Fig. 5a). Accordingly, the EP flux vectors mostly point downwards in the regions where $\hat{c}_g^z$ is positive and upwards in the regions where $\hat{c}_g^z$ is negative. For $n_x = 1$ retrograde waves, the upward and downward vertical EMF predominantly occur between 10 Pa and 1000 Pa in the northern mid-latitudes (Fig. 5b). Likewise, the EP flux vectors mostly point downwards or upwards in the regions where $\hat{c}_g^z$ is negative or positive respectively. Similar correlation between the orientations of EP flux vectors and the vertical EMF also presents during $L_s \sim 191°$ transfer events (Fig. 5c and 5d). In the regions poleward of 60° N and 60° S, the correlation breaks down for both transfer events perhaps due to a few reasons. First, the hydrostatic approximation that the EP flux theorem is based on is no longer appropriate when the scales of horizontal and vertical motions become comparable towards the poles. Second, the Fourier filter, which suppresses the numerical instabilities caused by the polar convergence of horizontal grids, acts as a diffusive wave energy sink that impacts the relation between the vertical EMF and wave energy flux.

In the low-to-mid latitudes between 60° S and 60° N , the vertical EMF may be used to identify possible wave sources since it is closely related to the wave energy flux. During $L_s \sim 261°$ transfer event, the positive $\overline{u'w'}$ from ground up indicates near-surface sources for $n_x = 1$ prograde waves at the southern low latitudes, while the negative $\overline{u'w'}$ at the northern low latitudes between 3000 Pa and the ground is likely generated by wave reflection or a local source near 3000 Pa (Fig. 5a). For $n_x = 1$ retrograde waves, the negative and positive $\overline{u'w'}$ (corresponding to the upward and downward wave energy flux respectively) near 40° − 50° N and at the equator between 30 Pa and 100 Pa indicate local wave sources since wave energy tends to radiate away from its source (Fig. 5b). During $L_s \sim 191°$ transfer event, the potential sources of $n_x = 1$ prograde waves still appear to be near the ground, but they locate at the northern low latitudes instead. The negative $\overline{u'w'}$ for $n_x = 1$ prograde waves in the southern hemisphere may be caused by wave reflection or a local wave source near 6000 Pa (Fig. 5c). For $n_x = 1$ retrograde waves, there appears a major wave source at the latitudes between 40° S and 60° S near 1000-2000 Pa (Fig. 5d). Both midlatitude wave sources, indicated by the wave energy flux associated with the $n_x = 1$ retrograde waves, coincide with the diverging EP flux vectors as well as those sources implied by the EP flux divergences in Fig. 3d and Fig. 4d. The probability of these wave sources will be examined using atmospheric stability criteria in Section 4.6.

We find that both the horizontal and vertical components of the EP flux are important in driving the zonal wind in our TitanWRF simulations. Figure 6 shows the horizontal and vertical divergences of total EP flux at selected pressure levels that correspond to the strongest fluxes. For $L_s \sim 261°$, the horizontal EMF dominates the jet acceleration at the equator. However, the vertical EP flux is not trivial since it can become just about a factor of two smaller than its horizontal counterpart at some pressure levels (Fig. 6a). The equatorial acceleration at 80 Pa and deceleration at 150 Pa due to the vertical component of EP flux (thin black lines in Fig. 6a) result from the vertical convergence of vertical EMF $\overline{u'w'}$, and they may be a result of absorption of upward propagating waves with positive intrinsic phase speed and downward propagating waves with negative intrinsic phase speed respectively since the zonal wind increases monotonically from 150 Pa to 80 Pa. Alternatively, they may a result of wave reflection near the critical layer where EP flux diverges (converges) above (below) the critical layer. The combination of horizontal and vertical divergence/convergence of EP fluxes leads to the net acceleration of the equatorial superrotation. At the northern mid-latitudes, the vertical divergence and the horizontal convergence have comparable strength, with the latter being slightly dominant, leading to slight net deceleration at these latitudes. For $L_s \sim 191°$, the horizontal EP flux dominates at most latitudes, but the contribution of the vertical EP flux is still noticeable at the southern mid-latitudes (Fig. 6b).

The EP fluxes in TitanWRF appear to be quite different from those in Lewis et al. [2023] in terms of their distribution and how they interact with the zonal winds during the similar time periods. For instance, the horizontal EP flux dominated almost all latitudes during the late northern autumn in Lewis et al. [2023], and it diverged at the southern low latitudes and converged at the northern low latitudes, whereas in our model the EP flux diverges at the equator and converges at the mid-to-high latitudes in the northern hemisphere. Moreover, the vertical EP flux suggests that waves propagate both upwards and downwards in our model (as permitted by the statically stable stratosphere), while they propagated mostly upwards in Lewis et al. [2023]. The difference may be caused by different wave source/sink locations predicted by the models. Despite the similar scaling on the horizontal and vertical EP flux components (see Fig. 3 captions for details), the vertical EP flux has a much stronger presence in our model compared to that in Lewis et al. [2023]. Note that the EP flux vectors in Newman et al. [2011] appear to be purely horizontal because they did not scale the horizontal and vertical components in their figure.

**4.1.2 Temporal variability of zonal acceleration**

The jet acceleration/deceleration shows different temporal behaviors during the two transfer events. Figures 7 and 8 show the zonal wind tendencies at selected pressure levels. During the $L_s \sim 261°$ transfer event, the equatorial zonal wind acceleration at lower altitude (higher pressure) lags behind that at higher altitude (lower pressure). For instance, the equatorial eastward acceleration peaks at $L_s = 260°$ for 50 Pa level, $L_s = 261°$ for 80 Pa level and $L_s = 262°$ for 150 Pa level. This temporal behavior does not present during the $L_s \sim 191°$ transfer event. Instead, the zonal wind is accelerated simultaneously across several pressure levels, although the zonal wind at higher altitude does experience slightly longer acceleration. Both the $L_s \sim 261°$ and $L_s \sim 191°$ transfer events show deceleration of zonal wind at mid-to-high latitudes, with the former in the northern hemisphere and the latter in the southern hemisphere. The acceleration near the equator and deceleration at mid-to-high latitudes may occur in several scenarios such as a tropical Rossby wave that dissipates at mid-to-high latitudes (*e.g.*, Vallis [2006]), or a result of R-K instability. The latter is discussed in sections 4.2 - 4.5 with evidence on the occurrence of such instability.

**4.2 Cospectral analysis**

**4.2.1 Global wave properties**

We perform cospectral analysis on the EMF to retrieve more information on the waves that are responsible for transferring eddy energy to the zonal flow. Such information includes wave phase speeds and their corresponding wavelengths (or wave frequencies), from which the types of planetary waves in the model results may be determined. Cospectral analysis on the horizontal EMF shows that the waves are prograde at low latitudes and retrograde at high latitudes relative to the zonal flow, and that low-wavenumber waves carry more energy than their high-wavenumber counterparts. There are few westward-propagating waves (westward phase speed relative to the planet rotation) compared to their eastward counterparts, so they are excluded from the results shown below.

Figure 9 shows the PSD of horizontal EMF as a function of latitude, phase speed, and zonal wavenumber at selected pressure levels where the wave activities are strongest during the $L_s$ ~ 261° transfer event. The analysis shows that multiple wave modes coexist, as indicated by the widespread PSD with varying phase speeds and zonal wavenumbers. The dominant wave modes (waves with the largest PSD) are mostly prograde at equatorial latitudes (*i.e.*, phase speed is eastward relative to the zonal flow) and retrograde (*i.e.*, phase speed is westward relative to the zonal flow) at mid-to-high latitudes. Overall, the PSD has maximum absolute values at the equator and rapidly decays towards mid-to-high latitudes, resembling the features of equatorially trapped planetary waves (Holton, [2004]). Moreover, the phase speed corresponding to the maximum of absolute PSD becomes closer to the zonal wind speed with increasing altitude (decreasing pressure) at the equator.

The properties of the wave packet are less complicated during the $L_s \sim 191°$ transfer event. Figures 10a, 10c and 10e show similar prograde wave modes at the equator and retrograde wave modes at mid-to-high latitudes, but the phase speed spectra corresponding to the maxima of absolute PSD has a smaller wavenumber range than the waves responsible for the $L_s \sim 261°$ event. This can also be seen in the zonal wavenumber distribution, which shows that wavenumber 1 or 2 modes dominate at most latitudes for all three pressure levels. Similar to $L_s \sim 261°$, the phase speed corresponding to the maximum of absolute PSD also becomes closer to the zonal wind speed with increasing altitude, and they become comparable at p=1170 Pa.

The comparison between the zonal mean zonal wind and the mean phase speed (constant angular velocity) in both transfer events suggests the likely existence of equatorially trapped Kelvin wave modes and mid-to-high latitude Rossby wave modes. We take the zonal wavenumber 1 geopotential anomaly and its corresponding horizontal wind perturbation for a close comparison with those demonstrated by Lewis et al. [2023]. Figure 11 shows the geopotential anomaly and eddy wind vectors at 80 Pa for the $L_s \sim 261°$ event and at 2000 Pa for the $L_s \sim 191°$ event. During both events, the eddy wind vectors are mostly pointing east-westward directions (correlated with peaks and troughs of geopotential anomalies) near the equator. These non-rotating, divergent motions are consistent with the characteristics of Kelvin waves. At northern high latitudes during $L_s \sim 261°$ event and southern high latitudes during $L_s \sim 191°$, the rotational motion of the wind vectors becomes strong, which exhibits eddy wind patterns similar to that of Rossby waves. Comparison of zonal wavenumber 1 geopotential anomaly and its associated wind perturbation between our results and those of Lewis et al. [2023] highlights the similarity in the wave patterns produced by both TitanWRF and TAM, which resemble coupled Rossby-Kelvin waves.

### 4.2.2 Zonal wind driven by planetary waves

The effect of prograde waves at low latitudes and retrograde waves at mid-to-high latitudes is to accelerate and decelerate the zonal winds at the corresponding latitudes when these waves dissipate locally. Figure 12a and 12b show the cospectra of the horizontal EMF convergence at the pressure levels of 80 Pa and 2000 Pa for $L_s \sim 261°$ and $L_s \sim 191°$ events respectively. The mean eddy acceleration (dashed lines) summed over all phase speeds and wavenumbers is mostly positive at the low latitudes and negative at the mid-to-high latitudes, the behavior is in good agreement with the divergence of EP flux (Fig. 3, 4 and 6).

The likelihood of spontaneous deceleration and acceleration by Rossby waves (mostly rotational) and Kelvin waves (mostly divergent) can be further examined using rotational and divergent components of the wind perturbations (Zurita-Gotor [2019]; Zurita-Gotor et al. [2022]). Figure 13 shows the zonally-averaged root-mean-squared geopotential anomaly as well as the divergent and rotational contributions to the EMF convergence by the dominant terms $\overline{u_r' D'}$ and $\overline{v_r' \zeta'}$ respectively for zonal wavenumber 1 (see Appendix C for definitions of these terms). The structures of the geopotential anomaly are similar to those described in Zurita-Gotor et al. [2022], which exhibit local maxima at both the equator and the mid-to-high latitudes. The equatorial maxima oscillating throughout the stratosphere indicate the presence of Kelvin waves, while the oscillating maxima at the mid-to-high latitudes are confined to higher altitudes and they indicate Rossby waves. These structures resemble the Rossby-Kelvin wave mode suggested by Zurita-Gotor et al. [2022] and Zurita-Gotor and Held [2018].

The divergent and rotational EMF convergence confirms the likelihood of zonal wind acceleration/deceleration by Rossby-Kelvin waves. For both transfer events, the divergent component produces equatorial acceleration (50-200 Pa for $L_s \sim 261°$ and 600-3000 Pa for $L_s \sim 191°$), while the rotational component produces mid-to-high latitude deceleration near the similar pressure levels (*e.g.*, in the regions poleward of 50° N for $L_s \sim 261°$ and poleward of 50° S for $L_s \sim 191°$). Since the acceleration and deceleration described here occur simultaneously but far from the critical lines (~ 40° N or S for the transfer events), the most logical explanation is that both Rossby waves and Kelvin waves become concurrently unstable and deposit their associated eddy momentum to the background flow, *i.e.*, the effect of R-K instability described in many literatures.

### 4.3 Equatorial wave modes

The cospectral analysis above roughly provides the zonal properties of the dominant wave modes that fit the characteristics of the planetary-scale waves. The zonal wavenumber 1 mode with maximum phase speed of about $200 ms^{-1}$ (wave period of about 1.5 Titan hours) is consistent with the dominant wave modes previously determined by Newman et al. [2011]. Here, we identify the types of the equatorial waves using the same method in Lewis et al. [2023], from which the jet driven mechanism at the equator may be determined (*e.g.*, Wang and Mitchell [2014]).

The commonly used method to determine equatorial wave types is based on linearized shallow water wave equations on a beta plane, with the assistance of a vertical structure equation (*e.g.*, Wheeler and Nguyen [2015]; Fedorov and Brown [2009]). In this method, prior knowledge of vertical wave modes corresponding to the dominant horizontal wave modes evaluated by cospectral analysis is needed to estimate an equivalent depth (*e.g.*, Kiladis et al., [2009]; Wheeler and Nguyen [2015]), which is then used to determine dispersion relations associated with various wave types (*e.g.*, Wheeler and Kiladis [1999]). Figure 14 shows the daily-averaged PSD (obtained via 1D FFT) of geopotential anomaly in atmospheric layers between selected vertical ranges $z_{bot}$ and $z_{top}$ as a function of vertical mode $m_z$ and latitude ϕ. For $L_s \sim 261°$ event, we choose $z_{bot} = 0$ km ($p \approx 1.44 \times 10^5 Pa$) and $z_{top} = 400\ km$ ($p \approx 0.25\ Pa$), with the latter being the effective model top that is outside of the region influenced by Rayleigh damping near the actual model top (~450 km). For $L_s \sim 191°$ event, we choose $z_{bot} = 0$ km and $z_{top} = 150\ km$ ($p \approx 200\ Pa$) since most wave-mean interactions are limited to the region deeper than a few hundred pascals (Figure 4d).

In general, the choices of the vertical domain for the 1D FFT analysis should not cause significant variations in the determination of dominant vertical wave modes as long as the waves do not experience significant absorption, damping or reflection in the domain. For instance, the vertical wavelengths of upward propagating waves can be shortened when approaching critical layer and stretched when being dissipated. For Titan, the eddy viscosity is not large enough to cause significant wave damping (*e.g.*, Li et al. [2013]). Additionally, Titan's atmosphere is mostly statically stable with a very shallow planetary boundary layer, which promotes the vertical propagation of planetary waves. As an experiment on the sensitivity of the selected vertical domain, we have compared the 1D FFT analysis for $L_s \sim 261°$ event using the combinations of $z_{bot} = 0\ km$, $z_{bot} = 100\ km$, $z_{top} = 200\ km$, and $z_{top} = 400\ km$ and found that the dominant wave modes are essentially the same in these cases. For $L_s \sim 191°$ event, however, the vertical wave modes in the vertical domain between $z_{bot} = 0\ km$ and $z_{top} = 100\ km$ are very different from those in the vertical domain starting from $z_{bot} = 100\ km$ (not shown). The only wave modes shared by these two vertical domains are the vertical wavenumber 1 at the mid-to-high latitudes. This is likely due to the critical layer absorption/reflection that prohibits the upward propagation of waves originating from the ground level (see Section 4.4). We choose the vertical domain between $z_{bot} = 0\ km$ and $z_{top} = 150\ km$ for $L_s \sim 191°$ just to include extra lower end of the spectrum (*e.g.*, to include waves with vertical lengths greater than 100 km).

Figure 14 shows that the PSD of geopotential anomaly is more pronounced at northern latitudes during the $L_s \sim 261°$ event, whereas it shows a relatively symmetric distribution with respect to the equator during the $L_s \sim 191°$ event, as expected by the seasonal variation of solar forcing (which becomes more symmetric in latitude when approaching equinox). For $L_s \sim 261°$ event, the dominant vertical wave mode is $m_z = 4$ near the equator, which corresponds to vertical wavelength $\lambda_z = \frac{z_{top} - z_{bot}}{m_z} \approx 100\ km$, while it is $m_z = 2\text{-}4$ at northern high latitudes and has a larger amplitude. For $L_s \sim 191°$ event, the prominent vertical wave modes near the equator have vertical wavenumber $m_z$ = 2-3, equivalent to vertical wavelength $\lambda_z = \frac{z_{top}-z_{bot}}{m_z} \approx 50 - 75 \, km$. At high latitudes (*i.e.*, 60° poleward), the $m_z = 1$ mode ($\lambda_z \approx 150 \, km$) presents in both hemispheres, but the southern hemisphere has more high-order wave modes ($m_z > 3$) than the northern hemisphere.

Some clarification on the uncertainties of the vertical wave modes is needed to properly interpret the wave types obtained via the FFT analysis. The vertical wave modes shown in Fig. 14 are considered as estimates (or mean vertical wave properties) because the FFT method assumes that each wave mode is invariant with the altitude. The reasons for uncertainties are: first, vertical wavelength varies in an atmosphere with variable background properties such as wind and temperature; second, vertical wavelength may vary with altitude due to damping such as eddy viscous dissipation or other numerical dissipation (*e.g.*, Vadas and Fritts [2005]). The variability of vertical wavelengths as a function of altitude will lead to some errors when determining the vertical wave modes. Therefore, we consider the vertical wave mode $m_z$ with a variability of $\pm 1$ as acceptable solutions.

Figure 15 shows the PSD of equatorial EMF convergence (obtained from the cospectral analysis) as a function of zonal wavenumber and intrinsic frequency. Here the positive (negative) EMF convergence represents zonal acceleration (deceleration). The black lines are the calculated dispersion relations (using the dominant vertical wavelengths $\lambda_z$ obtained above) for the $n = 1$ Rossby wave mode, $n = 0$ mixed Rossby-gravity wave mode (for negative zonal wavenumber) or eastward inertio-gravity wave mode (for positive zonal wavenumber), $n = -1$ Kelvin wave mode, and $n = 1$ inertio-gravity wave mode, where $n$ is the meridional mode number. Wheeler and Nguyen [2015] provided detailed descriptions for these equatorial waves including the theoretical considerations of their dispersion relations. Similar to Lewis et al. [2023], we use an effective Coriolis parameter $f_e = \frac{\overline{u}tan\phi}{r} + f$ to calculate the dispersion relations since the equatorial region in Titan's stratosphere is in cyclostrophic balance, where the centrifugal force is more important than the Coriolis acceleration (see Appendix D). Also note that in their formulation the wave frequency may be considered as intrinsic wave frequency when there is a constant background wind speed $\overline{U}$, i.e., $\frac{\partial}{\partial t} + \overline{U}\frac{\partial}{\partial x} = i(\omega - k_x\overline{U}) = i\hat{\omega}$ ). The pressure levels and the latitudes (p = 80 Pa and $\phi = 7.5°$ S for $L_s \sim 261°$ event, and p = 2000 Pa and $\phi = 7.5°$ N for $L_s \sim 191°$ event) are chosen because these locations exhibit the strongest meridional convergence of EMF during these transfer events (Fig. 3d, Fig. 4d and Fig. 12).

During the $L_s \sim 261°$ transfer event, the Kelvin wave with the zonal wavenumber $n_x = 1$ dominates the equatorial acceleration (Fig. 15a). The largest positive EMF convergence at $n_x = 1$ and $\hat{\omega} \approx 2 \times 10^{-5} s^{-1}$ correlates well with the equatorial Kelvin wave mode calculated by the dispersion relation using vertical wavelength $\lambda_z = 80\ km$ ($m_z = 5$, the thin solid black line in Fig. 15a). The positive maximum near $n_x = 2$ and $\hat{\omega} \approx 5.5 \times 10^{-5} s^{-1}$ corresponds to the equatorial Kelvin wave mode calculated with $\lambda_z = 100\ km$ ($m_z = 4$, the thick solid black line in Fig. 15a), which is mostly canceled by a negative maximum near $n_x = 3$ and $\hat{\omega} \approx 5.5 \times 10^{-5} s^{-1}$. The mixed Rossby-gravity wave intercepts with a pair of positive and negative maxima near $n_x = -3$ and $\hat{\omega} = 2 \times 10^{-5} s^{-1}$, while the n=1 Rossby wave comes across the positive EMF convergence near $n_x = -2\ or\ -3$ and $\hat{\omega} = 1 \times 10^{-5} s^{-1}$.

The zonal wavenumber $n_x = 1$ Kelvin wave again dominates the equatorial acceleration for the $L_s \sim 191°$ transfer event. Figure 15b shows the equatorial wave types determined by the vertical wave mode $\lambda_z = 50\,km$. The zonal wavenumbers 1, 2 and 3 all appear to contribute to the zonal acceleration but $n_x = 1$ Kelvin wave is more prominent. Unlike the $L_s \sim 261°$ case, no other wave types are observed.

The overall equatorial wave modes share some similarities with those in Lewis et al. [2023], but also have some differences. Lewis et al. [2023] showed that the equatorial acceleration was primarily caused by the wave modes with small intrinsic phase speeds at very low zonal wavenumbers ($n_x = 1$ or 2). While the wave modes responsible for the equatorial acceleration are also dominated by zonal wavenumber 1 for both transfer events in TitanWRF, the intrinsic phase speeds are relatively larger than those in Lewis et al. [2023] (e.g., $c_p - \bar{u} \sim 50\,ms^{-1}$ for $L_s \sim 261°$ event and $\sim 25\,ms^{-1}$ for $L_s \sim 191°$ event in our simulations). In fact, after examining the equatorial wave modes at various pressure levels in TitanWRF, we find that the equatorial acceleration by the zonal wavenumber 1 Kelvin wave is relatively insensitive to the intrinsic phase speed. This is also evident from Fig. 6a and 6b, where the divergence of horizontal EP flux at the equator remains comparable across several pressure levels, despite significant variations in intrinsic phase speed with pressure (Fig. 9 and Fig. 10).

The comparison between the cospectral analysis of equatorial EMF convergence and the theoretically calculated dispersion relations merely shows the presence of possible wave modes. It does not explain how waves and mean flow interact. Using $L_s \sim 261°$ event as an example, a prograde Kelvin wave near $n_x = 3$ and $\hat{\omega} \approx 5.5 \times 10^{-5}\,s^{-1}$ corresponds to a negative maximum (deceleration), while the zonal wavenumber $n_x = 1$ Kelvin wave corresponds to a positive maximum (acceleration). Similarly, the retrograde waves can either accelerate or decelerate the zonal flow (Fig. 15a). The possible reasons for the dichotomy are: 1. the cospectral analysis is performed on the horizontal component of the waves thus the contribution of the vertical component is ignored; 2. equatorial acceleration or deceleration depends on whether these waves are dissipated, or they are leaving their source regions (*e.g.*, Vallis [2006]).

**4.4 Vertical propagation of equatorial waves in TitanWRF and TAM**

The vertical propagation of the equatorial waves is mostly hindered by critical layers due to wave absorption/reflection and/or Rossby-Kelvin instability. Figure 16 shows the density scaled PSD of the horizontal EMF with zonal mean zonal wind overlaid, as a demonstration of wave-mean interactions near the equator. We show zonal wavenumber $n_x = 1$ only since using all zonal wavenumbers makes no notable changes in our analysis. The overall PSD looks simpler compared to those in Lewis et al. [2023]'s analysis, which showed more wave activities perhaps due to including methane moist convection in their simulation. At $L_s \sim 261°$ (comparable to the solsticial analysis in Lewis et al. [2023]), two $n_x = 1$ wave modes with phase speeds of about $25\ ms^{-1}$ and $190\ ms^{-1}$ are generated in the lower atmosphere, with the faster one being the equatorial Kelvin wave. The slower $25\ ms^{-1}$ wave mode extends from the ground to a critical layer right below the tropopause at $1 \times 10^4$ Pa (0.1 bars), above which it shows little presence. This is consistent with the behavior of critical-layer wave absorption. The faster $190\ ms^{-1}$ wave mode experiences some damping effect in the lowest model layers, likely due to diffusive mixing in the thin PBL that extends merely a few hundred meters from the ground (*e.g.*, Tokano et al. [2006]), then becomes strong above the tropopause. It continues to propagate upwards until it reaches about 50 Pa, above which it becomes very weak largely due to wave dissipation in the vicinity of 50 Pa – 150 Pa, where the R-K instability leads to the deposition of equatorial Kelvin wave energy into the zonal wind. The wave absorption and reflection by critical layers also weaken the vertical propagation of the equatorial Kelvin wave, as indicated by the EP flux vectors and vertical convergence of vertical EMF (Fig. 3d and Fig. 6a). The critical-layer wave absorption may also explain the downward migration of near constant peak wind speed ($\sim 190 \text{m} s^{-1}$) over time (Fig. 16a) (see driving mechanism of the QBO in Earth's stratosphere for example, *e.g.*, Vallis [2017]).

The vertical propagation of equatorial waves and their interaction with the zonal wind near $L_s \sim 191°$ look very different to that during $L_s \sim 261°$ transfer event. At $L_s \sim 191°$, the $n_x = 1$ Kelvin wave with a single phase speed of about $130 m s^{-1}$ originates from the ground level and is damped within the thin PBL. The wave amplitude continues to decrease from $1 \times 10^5$ Pa (1 bar) to $2 \times 10^4$ Pa (0.2 bars), as shown by very small EMF in the region (Fig. 16b). The wave damping in this region may result from destructive interference between upward and downward propagating waves in the equatorial region, as indicated by the downward EP flux vectors (upward wave energy flux) on the northern side and the upward EP flux vectors (downward wave energy flux) on the southern side of the equator, respectively (Fig. 5c). The wave continues to travel upwards until it reaches the pressure level p=1170 Pa, above which the wave breaking/dissipation due to R-K instability and the critical layer absorption/reflection stop further propagation of the wave. Unlike the time evolution of the zonal wind profile at $L_s \sim 261°$, the zonal wind profiles do not create descending critical layers during the $L_s \sim 191°$ transfer event but the zonal wind increases in the region between 1000 Pa and 3000 Pa during the event. This suggests that the equatorial wind acceleration in this region is likely due to R-K instability, which is consistent with the behavior of the EP fluxes near equator, *i.e.*, the horizontal divergence of horizontal EP flux is much larger than the vertical divergence of vertical EP fluxes (Fig. 6b).

The critical layers act as a major barrier for wave propagation but waves may be absorbed and reflected even before they reach these layers. Figure 17 shows the refractive index $\frac{f^2}{N^2} n_{ref}^2$ with EP flux vectors overlaid. The critical layers form distinct boundaries that separate maxima and minima of refractive index, as expected by the relation $n_{ref}^2 \propto \frac{1}{c-\bar{u}}$. As depicted by the EP flux vectors, the equatorial waves originating from the ground level propagate upwards in the region with weakly positive $n_{ref}^2$, get refracted once they reach the vicinity of reflecting surface ($n_{ref}^2 \sim 0$, about 300-400 Pa for $L_s \sim 261°$ and 3000-4000 Pa for $L_s \sim 191°$ at the equator), and cease to propagate in strong wave evanescent regions (corresponding to the regions with large negative $n_{ref}^2$). Waves propagating above the wave evanescent region likely originate from local sources, such as those shown by the upward-downward EP fluxes near 50 Pa at the northern midlatitudes for $L_s \sim 261°$ (Fig. 17a) and near 2000 Pa at the southern midlatitudes for $L_s \sim 191°$ (Fig. 17b).

The relative magnitude of the upward and downward EP flux vectors shows that the equatorial waves experience nearly complete reflection for $L_s \sim 191°$, suggesting these waves are trapped between the reflecting surface and the ground (Fig. 17b). For $L_s \sim 261°$, the equatorial waves experience more absorption than reflection near the reflecting surface since the downward EP flux is significantly weaker than the upward EP flux (Fig. 17a). During both transfer events, the absorption and reflection of upward propagating waves occur in the region deeper than the critical layers despite Titan's stratosphere is convectively stable (*i.e.*, the bulk Richardson number $R_i \gg 1$). Therefore, the wave evanescent regions sandwiched between the reflecting surface and the critical layers are likely the results of wave breaking due to R-K instability and wave propagation into the horizontal critical layers.

The behavior of the equatorial waves appears to be somewhat different from that in Lewis et al. [2023]. Here, we use the wave phase speed spectra at $L_s \sim 261°$ in our simulation as a comparison to the similar analysis at $L_s = 226° - 234°$ in Lewis et al. [2023]. For the slow-traveling wave with $c_p = 25 \text{ m}s^{-1}$ in our simulation, a similar wave mode also existed in Lewis et al. [2023] (as shown in their Figure 9) but was able to travel beyond the critical layer present in their simulation until the wave reached a few hundred Pa. For the fast-traveling wave with $c_p = 190 \text{ m}s^{-1}$ in our model, a similar wave mode did not seem to exist in Lewis et al. [2023]. Instead, they showed that prograde waves with $c_p > 100 \text{ m}s^{-1}$ occurred for a wide range of phase speeds and pressures in the lower stratosphere (see Fig. 9 in Lewis et al. [2023]). Moreover, Lewis et al. [2023] mentioned that the eddy acceleration was almost entirely due to horizontal divergence (as the scaled EP flux vectors appeared to be mostly horizontal at low latitudes in their model results). It is unclear whether absorption of vertically propagating Kelvin wave would occur in their simulation.

**4.5 Resonance condition for Rossby-Kelvin instability**

The resonant coupling between Rossby and Kelvin waves can effectively drive equatorial superrotation in planetary atmospheres (including both slowly rotating or Earth-like planets, e.g., Iga and Matsuda [2005]). The resonance condition, as shown by Wang and Mitchell [2014], requires that the ratio (*i.e.*, the Froude number $F_r$ defined in Section 3.2.2) between the Doppler-shifted frequency of mid-to-high latitude Rossby waves and that of the equatorial Kelvin waves approaches unity. Iga and Matsuda [2005] depicted the resonant scenario in which the equatorial Kelvin wave grows via supply of angular momentum emitted by the Rossby wave from higher latitudes, and thus both waves continue to grow, since they provide opposite angular momentum (opposite intrinsic phase speed) to each other (see Fig. 2 in Zurita-Gotor and Held [2018] for an illustration). Here we examine whether such condition is met during the transfer events in TitanWRF.

Figure 18 shows the Froude number calculated for the $n_x = 1$ Rossby and Kelvin wave modes at $L_s = 261.5°$ and $L_s = 192.3°$, shown as a function of pressure over the vertical extent of interest (10 - $10^4$ Pa). The Froude number increases with increasing altitude (decreasing pressure), and reaches the theoretical resonant condition $F_r = 1$ near 800 Pa at $L_s = 261.5°$ and 3000 Pa at $L_s = 192.3°$. Overall, the Froude number is closer to unity at $L_s = 261.5°$ than at $L_s = 192.3°$ for pressure levels above 1000 Pa, and for both cases it lies well within the range of values (1-3) for R-K instability identified via linear wave analysis by Wang and Mitchell [2014]. However, $F_r \sim 1$ does not always mean equatorial Kelvin waves will break and dissipate. For instance, the equatorial wind shows little change between 300 Pa and 800 Pa during the $L_s \sim 261°$ transfer event (Fig. 16a) despite the Froude number is closest to unity (hence the resonance should be greatest, according to theory). Similarly, Lewis et al. [2023] noted that such criterion did not guarantee the occurrence of EMF convergence. Therefore, the criterion $F_r = 1$ may be necessary but not sufficient for the occurrence of Rossby-Kelvin wave resonance.

**4.6 Sources of Rossby and Kelvin waves**

The generation of atmospheric waves, such as Rossby and Kelvin waves, requires disturbances that destabilize the atmosphere. These disturbances are often associated with barotropic and baroclinic instabilities (see section 3.2.1). In TitanWRF, we find that the planetary scale waves are generated by a combination of baroclinic and barotropic instabilities, by contrast with the earlier analysis by Newman et al. [2011] which originally identified barotropic instabilities as the main source of waves at low-to-mid latitudes, as they only considered the possible wave generation in the stratosphere. Likewise, Lewis et al. [2023] also suggested that the planetary waves possibly have barotropic sources in the stratosphere using a more generalized barotropic stability criterion. However, both studies omitted the fact that the stratosphere at low latitudes is ageostrophic given the large zonal wind speed and small Coriolis parameter. Figure 19 shows the zonal mean quantities associated with the meridional wind curvature (2D barotropic stability criterion; section 3.2.1) and meridional IPV gradient (Charney-Stern 3D barotropic stability criterion; section 3.2.1), and inertial stability criterion (section 3.2.3) near 80 Pa at $L_s = 261.5°$. The violation of the 2D barotropic stability criterion occurs mostly at low-to-mid latitudes, where the zonal wind curvature exceeds β. Similarly, the Charney-Stern barotropic stability criterion is violated at a few similar latitudes where the IPV gradient changes sign, *e.g.*, near ±10°. These violations in the stratosphere, however, may not hold in the equatorial region, where a small Rossby number ($R_o \ll 1$) is required for the QG approximation to be valid. No apparent violation of inertial stability is found as $f(\zeta + f)$ remains positive at all latitudes, except it dips very close to zero in the vicinity of the equator, where local wave generation may not be ruled out since the atmosphere is marginally stable with respect to inertial stability. We evaluated these stability criteria for $L_s \sim 191°$ transfer event and found very similar behavior (not shown).

We next examine the possible occurrences of baroclinic instability in two major regions: the interior of the atmosphere (*i.e.,* troposphere and stratosphere) and/or the lowest layer in our model (see section 3.2.1 for the two necessary but not sufficient conditions of baroclinic instability). Figures 20 and 21 show the spatial distribution of meridional IPV gradient as a function of latitude and pressure, as well as the relationship between the QGPV gradient and the vertical wind shear in the lowest model layer. The former is for examining the baroclinic instability in the interior of the atmosphere and the latter is for examining the baroclinic instability in the lowest atmospheric layer. During both the $L_s \sim 261°$ and $L_s \sim 191°$ transfer events, the IPV gradient does not change sign in the interior of the atmosphere, with a few exceptions including some regions above 200-300 Pa and in the north polar regions (Fig 19a and 20a). Baroclinic instability may occur in these regions.

The violation of baroclinic stability criterion in the lowest model layer (the atmospheric layer right above the ground) suggests that the equatorial Kelvin waves may have baroclinic sources near the ground. The QGPV gradient and the vertical shear of the zonal wind have the same sign in the equatorial region (about $\pm 25°$) for $L_s \sim 261°$ event, and similarly $\pm 10°$ for $L_s \sim 191°$ event (Fig 20b and 21b). The violation of the baroclinic stability criterion near the ground at the equator is consistent with the vertical distribution of the EMF PSD shown in Fig. 16, which suggests that the dominant waves originate somewhere near the ground. Moreover, these potentially baroclinically unstable locations partially correlate with the distributions of upward-propagating equatorial Kelvin waves indicated by the upward wave energy fluxes (*i.e.*, in the latitudes between the equator and 20° S for $L_s \sim 261°$ and between the equator and 50° N for $L_s \sim 191°$, see Fig. 5a and Fig. 5c) as well as the oscillating patterns of geopotential anomalies in the vertical direction (Fig. 13).

The midlatitude Rossby waves during both transfer events, on the other hand, are unlikely a result of baroclinic instability despite the violation of stability criterion at the northern mid-to-high latitudes for $L_s \sim 261°$ and at both northern and southern midlatitudes for $L_s \sim 191°$. Neither the EP flux vectors nor the vertical profiles of geopotential anomaly suggests near-surface sources of the midlatitude Rossby waves. Instead, they show likely Rossby waves sources in the region between 10 Pa and 100 Pa for $L_s \sim 261°$ and between 1000 Pa and 3000 Pa for $L_s \sim 191°$ at the latitudes where 2D and/or 3D barotropic stability criterions are violated.

## 5. Conclusion and discussion

We perform an analysis of TitanWRF simulation results equivalent to those in Newman et al. [2011] in order to better understand the mechanisms that maintain the stratospheric superrotation. We focus on the wave-mean interactions as well as the generation of the planetary waves that are responsible for jet acceleration/deceleration during the strongest 'solsticial' and 'equinoctial' angular momentum transfer events, at $L_s \sim 261°$ and $L_s \sim 191°$. Similar to Newman et al. [2011] and recent studies by Lewis et al. [2023], we find the dominant wave modes that contribute eddy momentum to the zonal flow are low-order zonal wave modes, with wavenumber 1 most prominent. More specifically, we confirm that, as found by Lewis et al. [2023], the R-K instability generated by the resonance of Rossby-Kelvin waves is likely the primary source of prograde angular momentum driving the equatorial superrotation.

Cospectral analysis of EMF reveals the relationships between wave phase speed and speed of the background zonal flow. During the $L_s \sim 261°$ transfer event, the dominant waves travel zonally with speeds comparable to or faster than the background zonal flow at the equator at pressure levels where there are clear signs of zonal wind acceleration, and are faster (slower) than the background flow at southern (northern) midlatitudes. During the $L_s \sim 191°$ transfer event, the dominant waves travel faster at the equator, and slower at mid-to-high latitudes, than the background flow. The change in the sign of the relative speed between waves and background flow (intrinsic phase speed) as a function of latitude creates critical layers at mid-latitudes, which is a typical occurrence when equatorial Kelvin waves are trapped by mid-latitude Rossby waves. The dominant wave modes during both transfer events are zonal wavenumber 1 and 2, consistent with previous analysis by Newman et al. [2011] and Lewis et al. [2023].

We determine the equatorial waves are primarily Kelvin waves with signatures of other waves, including mixed Rossby-gravity waves and inertia gravity waves, by comparing the PSD distribution in wave frequency-wavenumber space and the theoretical dispersion relations of various equatorial wave modes. The equatorial wave modes are very different between the two transfer events. The $L_s \sim 191°$ event shows distinct Kelvin waves with little presence of other wave types. The $L_s \sim 261°$ event, on the other hand, shows a mixture of Kelvin waves, Rossby waves and mixed Rossby-gravity waves. During both transfer events, the zonal wavenumber 1 equatorial Kelvin waves are more powerful than other wave modes.

The driving mechanisms of the equatorial superrotation due to wave-mean interactions come in two variations in TitanWRF. First, eddies produced by R-K instabilities, due to the resonance of the two wave modes, deposit their eastward momentum near the equator and westward momentum at midlatitudes, leading to acceleration of the equatorial zonal flow and deceleration of mid-latitude zonal flow. This mechanism occurs for both $L_s \sim 261°$ and $L_s \sim 191°$ transfer events, as shown by several aspects of wave-mean interactions including spontaneous equatorial acceleration and mid-latitude deceleration of zonal wind by the phase locking equatorial Kelvin wave and mid-latitude Rossby wave (corresponding to Froude number between 1 and 2) in our cospectra analysis, as well as the divergent (rotational) contribution to the EMF convergence that leads to the zonal acceleration (deceleration) near the equator (midlatitudes). Second, there is equatorial acceleration due to the absorption of vertically propagating Kelvin waves near the critical layers, but it is weaker than the contribution of R-K instability. This occurs in the region above 100 Pa for $L_s \sim 261°$ and near 1000 Pa for $L_s \sim 191°$, with the wave absorption during $L_s \sim 261°$ event being more noticeable given the larger ratio between the vertical and horizontal components of the EP flux divergence and the smaller ratio between the downward (reflected waves) and upward (incident waves) EP flux magnitudes.

Lewis et al. [2023] acknowledged that a critical region also existed in their model results but suggested that its role was to prohibit the vertical propagation of waves only, vs also leading to wave absorption. The difference between our results and those in Lewis et al. [2023] may be caused by the numerical representations of nonlinear processes, such as numerical and spatial discretization and choice of dynamical solver, since these two models have different dynamical cores (*e.g.*, TitanWRF is a finite difference model whereas TAM is a spectral model), and the physics processes are also somewhat different (*e.g.*, the version of TitanWRF used here is a dry model, while Lewis et al. [2023]'s model includes a methane hydrological cycle).

We further examine the sources of the Kelvin and Rossby waves and find that the equatorial Kelvin waves likely have baroclinic sources near the ground, while the mid-latitude Rossby waves are likely generated by barotropic instabilities. In our simulation, barotropic instabilities (both in 2D and 3D regimes) may occur in the stratosphere at low-to-high latitudes but are not present in the equatorial region where EMF is strongest (*e.g.*, $-10°$ to $5°$ for $L_s = 261°$). Baroclinic instabilities, on the other hand, may occur in the lower troposphere both in the equatorial region and at the mid-to-high latitudes. Since the QG approximation at low latitudes is possibly valid only in the lower troposphere where wind speed is small, the equatorial Kelvin waves are more likely generated by baroclinic sources near the ground and propagate into the stratosphere, as also indicated by the upward wave energy flux, the oscillation patterns in geopotential anomaly and the vertical distribution of the PSD of EMF near the equator. The mid-to-high latitude Rossby waves, consistent with the findings in Mitchell et al. [2016], may have mixed barotropic-baroclinic sources in the stratosphere since the geopotential anomaly does not exhibit oscillation patterns in the lower stratosphere at the latitudes where the vertical EMF divergence is strongest (*i.e.*, $50°$ N to $75°$ N for $L_s = 261°$ and $50°$ S to $75°$ S for $L_s = 191°$). Note that the model results used for our analysis are from a dry atmosphere. A reanalysis of wave sources may be needed when hydrological cycle of methane is considered, since the latent heating and cooling associated with moist convection can impact the development and growth of baroclinic instabilities (*e.g.*, Banon [1986]; Emanuel et al. [1987]).

In this study, we only focus on the maintenance of equatorial superrotation due to wave-mean interactions during equinoctial and solsticial angular momentum transfer events in a spun-up model atmosphere. There are Kelvin/inertia-gravity wave activities during the gap periods but they are several orders of magnitude weaker than those during the transfer events, therefore they are excluded from this study. However, additional work is required to fully understand the role of atmospheric waves on the dynamics. For instance, Newman et al. [2011] noted that the stratospheric superrotation might be driven by the meridional propagation of planetary scale waves, due to absorption of those waves at critical layers. This scenario may occur at mid latitudes where critical layers are located. Nonetheless, the equator is far away from the critical lines and may be beyond the reach of the critical-layer effect for the two transfer events we analyzed. Additional analysis of other angular momentum transfer events is needed to thoroughly understand the influence of critical layers on the equatorial superrotation.

There are other factors that will affect our current understanding of wave-mean interactions in Titan's atmosphere. Besides the lack of moist convection in this study, the inclusion of topography will introduce orographic waves that are stationary with respect to the ground. Such waves will exert a drag on wind when they propagate vertically and break at the altitude where the wave amplitudes become sufficiently large, due to decreasing density with increasing altitude (*e.g.*, waves can become convectively unstable and break when the temperature perturbation is large enough). Another feature that is likely related to wave-mean flow interactions, yet is not captured by either the TitanWRF or TAM simulations, is the existence of a wind minimum near 75 km above the ground, which was observed by the Huygens probe during its entry (Folkner et al. [2006]). This wind minimum may be a result of wave-mean interactions between Rossby waves and/or orographic waves and the zonal flow, which shall be studied in the future. Moreover, the TAM simulations showed eddy angular momentum flux variability on time scale of a few $L_s$ but the magnitudes of these variabilities were much less than those during the angular momentum transfer events in TitanWRF. A future inter-model comparison will be needed to determine the detailed differences of wave generation and propagation and wave-mean interaction between the models.

## 6. Appendix

### A. Transformed Eulerian Mean equations and EP fluxes

The equations describing the zonal acceleration due to eddy-mean interactions are usually in form of the transformed Eulerian mean (TEM) equations. We provide the equations here again for convenience, with the same terms 1-5 defined as in Newman et al. [2011]. Following Andrews et al. [1987], the zonal momentum equation is:

$$\frac{\partial \overline{u}}{\partial t} = -\overline{v^*}\left(\frac{1}{a\cos\phi}\frac{\partial \overline{u}\cos\phi}{\partial \phi} - f\right) - \overline{w^*}\frac{\partial \overline{u}}{\partial z} + \frac{1}{\rho_0 a\cos\phi}\nabla \cdot F + \overline{X} \quad (A1)$$

Where $f$ is the Coriolis parameter, $\rho_0$ is the reference density (zonal mean here), $z$ is the height and the overbar $\overline{X}$ is the zonal mean of any physics quantity $X$. The left-hand side is the total acceleration on the zonal wind (term 1). The first term and the second term on the right-hand side (RHS) are the zonal acceleration due to meridional (term 2) and vertical (term 3) components of the residual circulation, where the residual mean velocities are defined as

$$\overline{v^*} = \overline{v} - \frac{1}{\rho_0}\frac{\partial}{\partial z}\left(\frac{\rho_0 \overline{v'\theta'}}{\partial \overline{\theta}/\partial z}\right) \quad (A2)$$

$$\overline{w^*} = \overline{w} + \frac{1}{a\cos\phi}\frac{\partial}{\partial \phi}\left(\frac{\overline{v'\theta'}\cos\phi}{\partial \overline{\theta}/\partial z}\right) \quad (A3)$$

Where $v$, $w$ and $\theta$ are the meridional velocity, zonal velocity and potential temperature respectively. The third term on the RHS of Eq. A1 is the contribution from EP flux (term 4), defined as:

$$\nabla \cdot F = \frac{1}{a\cos\phi}\frac{\partial}{\partial \phi}(F_\phi \cos\phi) + \frac{\partial F_z}{\partial z} \tag{A4}$$

With

$$F_\phi = \rho_0 a\cos\phi \left(\overline{v'\theta'}\frac{\partial \bar{u}/\partial z}{\partial \bar{\theta}/\partial z} - \overline{u'v'}\right) \tag{A5}$$

$$F_z = \rho_0 a\cos\phi \left[\left(f - \frac{1}{a\cos\phi}\frac{\partial \bar{u}\cos\phi}{\partial \phi}\right)\frac{\overline{v'\theta'}}{\partial \bar{\theta}/\partial z} - \overline{u'w'}\right] \tag{A6}$$

The vertical EMF $\overline{u'w'}$ in Eq. A6 is usually small compared to the eddy heat flux for hydrostatically balanced atmosphere, but we find that it cannot be ignored particularly in the lower stratosphere (see Section 4.1.1). The last term (term 5) on the RHS of Eq. A1 is the physical and/or numerical dissipation that may be represented by viscous turbulent mixing, *e.g.*, $\overline{X} \sim \nu\nabla^2 u$, where $\nu$ is eddy viscosity that is normally parameterized in atmospheric models (Holton, [2004]). This term is ignored in our discussion for simplicity.

**B. Cospectral analysis**

The cospectra (*i.e.*, space-time cross-spectral) analysis follows the same method described in Rand and Held [1991], which utilizes the time series of two dynamical variables such as wind, temperature or geopotential anomaly along longitude. For any two longitude-time series of variables $a(\lambda, t)$ and $b(\lambda, t)$, space-time Fourier transformation is performed on their perturbations with respect to either zonal mean (stationary eddies) or time average (transient eddies). The space-time Fourier transformation (using $a(\lambda, t)$ as an example) is defined as (Eq. 1b in Rand and Held [1991]):

$$A_{n,\omega} = \frac{1}{2\pi T} \int_0^T dt \int_0^{2\pi} a(\lambda, t) e^{-i(n\lambda + \omega t)} d\lambda \quad (B1)$$

Where $t$ is the time in seconds, $\lambda$ is the longitude in radians, $T$ is the duration of the time series (1 Titan day typical), $n$ is the number of waves that may fit around a zonal circle (or a segment of a circle) at a given latitude, and $\omega$ is the wave frequency with positive and negative values for eastward and westward propagating phases respectively. In discrete Fourier transformation, $n$ has a upper limit of $N/2$ where $N$ is the number of grid points in a zonal circle at a given latitude. $\omega$ is within the Nyquist limit $\pm\frac{1}{2\Delta t}$, where $\Delta t$ is the sampling frequency in seconds (10 Titan minutes typical without causing aliasing for zonal wavenumber 1 traveling at about $200 \text{ms}^{-1}$).

The cospectra power density in wavenumber-frequency space can be defined as (Eq. 1a in Rand and Held [1991]):

$$K_{n,\omega}(a, b) = 2 < \text{Re}(A_{n,\omega} B^*_{n,\omega}) > \quad (B2)$$

Where $B_{n,\omega}$ is the space-time Fourier transformation of $b(\lambda, t)$, $B^*_{n,\omega}$ is the conjugate of $B_{n,\omega}$, $Re$ means the real part, and the angled bracket <> means the average over a frequency bandwidth ($\Delta t$ in our calculations). To obtain the power spectral density in wavenumber-phase speed space, a conversion can be performed as (Eq. 3a in Rand and Held [1991]):

$$K_{n,c} \cdot \Delta c = K_{n,\omega} \cdot \Delta\omega \quad (B3)$$

Where $\omega = c\frac{n}{a\cos\phi}$, $a$ is the planet radius, $c$ is the phase speed and $\phi$ is the latitude. The maxima of $K_{n,\omega}$ or $K_{n,c}$ correspond to the dominant wave modes in wavenumber-frequency space or wavenumber-phase speed space.

## C. Rotational and divergent contributions to the EMF convergence

We use the same method in Zurita-Gotor and Held [2018] to decompose the convergence of horizontal EMF to obtain its rotational and divergent contributions from horizontal winds in spherical coordinates (Eq. 8 in Zurita-Gotor and Held [2018]):

$$-\frac{1}{a\cos^2\phi}\frac{\partial \overline{u'v'}\cos^2\phi}{\partial \phi} = \overline{v'\zeta'} - \overline{u'D'} \qquad (C1)$$

where the overbar is the zonal mean, $u'$ and $v'$ are the zonal and meridional wind perturbations with respect to the zonal mean. $\xi$ is the vorticity and $D$ is the divergence. The horizontal wind $\boldsymbol{v} = u\boldsymbol{i} + v\boldsymbol{j}$ can be decomposed to its rotational and divergent components (see Trenberth and Chen [1988], Hammond and Lewis [2021] for examples):

$$\boldsymbol{v} = \boldsymbol{v_r} + \boldsymbol{v_d} \qquad (C2)$$

In the above equation, the rotational wind component $\boldsymbol{v_r} = \boldsymbol{k} \times \nabla\psi$ and the divergent wind component $\boldsymbol{v_d} = \nabla\chi$, where ψ is the horizontal streamfunction and χ is the velocity potential. The dominant rotational and divergent contributions to the EMF convergence are $\overline{v'_r\zeta'}$ and $\overline{u'_rD'}$ respectively according to *e.g.*, Zurita-Gotor et al., [2019]. We found the same in our Titan model results, *e.g.*, the rotationally coupled divergent contribution to the EMF convergence is far greater than the pure divergent contribution ($\left|\overline{u'_rD'}\right| \gg \left|\overline{u'_dD'}\right|$) near the equator, which implies that the coupling of Rossby and Kelvin waves contributes more to the equatorial acceleration than pure Kelvin waves since Rossby waves are mostly rotational while Kelvin waves are mostly divergent.

### D. Effective Coriolis parameter

Slow-rotating planets such as Titan and Venus have very weak Coriolis effect particularly at low latitudes. Thus, the Coriolis effect may be superseded by other rotational effect such as the curvature (also known as metric) terms in the governing equations of atmospheric circulations (*e.g.*, Holton [2004]). These terms are often ignored in linear wave theories since they are very small for geostrophic-balanced flow. In the horizontal momentum equations, the curvature terms for the zonal and meridional winds are $\frac{uv\tan\phi}{a}$ and $\frac{u^2\tan\phi}{a}$ respectively, where $u$ and $v$ are zonal and meridional winds, $\phi$ is the latitude and $a$ is the planet radius. They can be combined with the Coriolis terms as $\left(\frac{u\tan\phi}{a} + f\right)v$ and $\left(\frac{u\tan\phi}{a} + f\right)u$. The effective Coriolis parameter is thus defined as $f_e = \frac{u\tan\phi}{a} + f$, which shall be used in place of Coriolis parameter $f$ when deriving dispersion relations for planetary waves. For the equatorial region, $f$ is assumed to vary linearly along latitude (β-plane approximation) so the wave equations can be simplified (Wheeler and Nguyen [2015]). Here $\beta = \frac{\partial f_e}{\partial y}$ and it can be expressed as $\beta \approx \frac{\overline{U}}{a^2} + \frac{2\Omega}{a}$ at the equator, where $\overline{U}$ is the mean equatorial zonal wind ( $\sim 160\, ms^{-1}$ at 80 Pa for $L_s = 261°$ and $\sim 90\, ms^{-1}$ at 2000 Pa for $L_s = 191°$ ) and $\Omega$ is the Titan's rotation rate $\sim 4.6 \times 10^{-6} s^{-1}$.

### E. Vertical wave propagation

We use the meridional gradient of QGPV with some characteristic wave properties to evaluate how equatorial waves propagate vertically. The method follows that in Harnik and Lindzen [2001] but we ignore meridional wave propagation for simplicity. The analysis is adequate for Kelvin waves since they do not have meridional structures. Harnik and Lindzen [2001] showed that the properties of vertical wave propagation could be characterized by a refractive index $n^2_{ref}$, defined as

$$n_{\text{ref}}^2 = \frac{N^2}{f^2}\left[-\frac{\partial_y \overline{q}_G}{c - \overline{u}} - k_x^2 + F(N^2)\right] \tag{E1}$$

where $q_G$ is the QGPV defined in Section 3.2.1, $\partial_y \overline{q}_G$ is the meridional gradient of $q_G$, $k_x$ is the zonal wavenumber, $F(N^2) \equiv f^2 \frac{e^{z/2H}}{N} \frac{\partial}{\partial z}\left[\frac{e^{-z/H}}{N^2} \frac{\partial}{\partial z}(e^{z/2H} N)\right]$, $N$ is the zonal mean Brunt Väisälä frequency and $H$ is the zonal mean pressure scale height. The phase speed $c = c_\circ \cos(\phi)$ is a function of latitude only, where $c_\circ$ is the typical wave phase speed at the equator, i.e., $190 ms^{-1}$ for $L_s \sim 261°$ and $130 ms^{-1}$ for $L_s \sim 191°$. In an atmosphere with slow-varying base state such as wind and temperature, $F(N^2)$ can be simplified to $F(N^2) = -\frac{f^2}{N^2}\frac{1}{4H^2}$. On the RHS of Eq. E1, the first term in the bracket dominates in this kind of atmosphere. Moreover, it is convenient to use $\frac{f^2}{N^2} n_{ref}^2$ instead of $n_{ref}^2$ to describe the properties of vertical wave propagation since $f$ approaches zero near the equator.

The refractive index $n_{ref}^2$, mathematically the vertical wavenumber squared (Harnik and Lindzen [2001]), can be used to describe whether vertically propagating waves can transmit or be absorbed and reflected. For positive $n_{ref}^2$, vertical propagation of waves is permitted. For negative $n_{ref}^2$, waves are evanescent, meaning wave amplitudes decay rapidly. These two regimes are separated by a turning surface where $n_{ref}^2 = 0$, which is also called reflecting surface where wave absorption and reflection likely occur. Equation E1 also shows that, for retrograde waves ($c - \overline{u} < 0$), positive $\partial_y \overline{q}_G$ is for wave propagation and negative $\partial_y \overline{q}_G$ is for wave absorption and reflection, and vice versa for prograde waves. Strong absorption and reflection of upward propagating waves are expected when the reflecting surface is near the critical lines since $n_{ref}^2 \propto \frac{1}{c - \overline{u}}$, and a strong wave evanescent region ($n_{ref}^2 \ll 0$) above the reflecting surface can effectively reflect or even over reflect the incident waves (*e.g.*, Lindzen and Tung [1976]).

**Acknowledgement**

The authors would like to thank the JPL Office of Research and Development Topical Research and Development Program for funding. The research was carried out at the Jet Propulsion Laboratory, California Institute of Technology, under a contract with the National Aeronautics and Space Administration (80NM0018D0004). This work was also supported by NASA Astrobiology Institute "Habitability of Hydrocarbon Worlds: Titan and Beyond" funding via JPL subcontract #1643481.

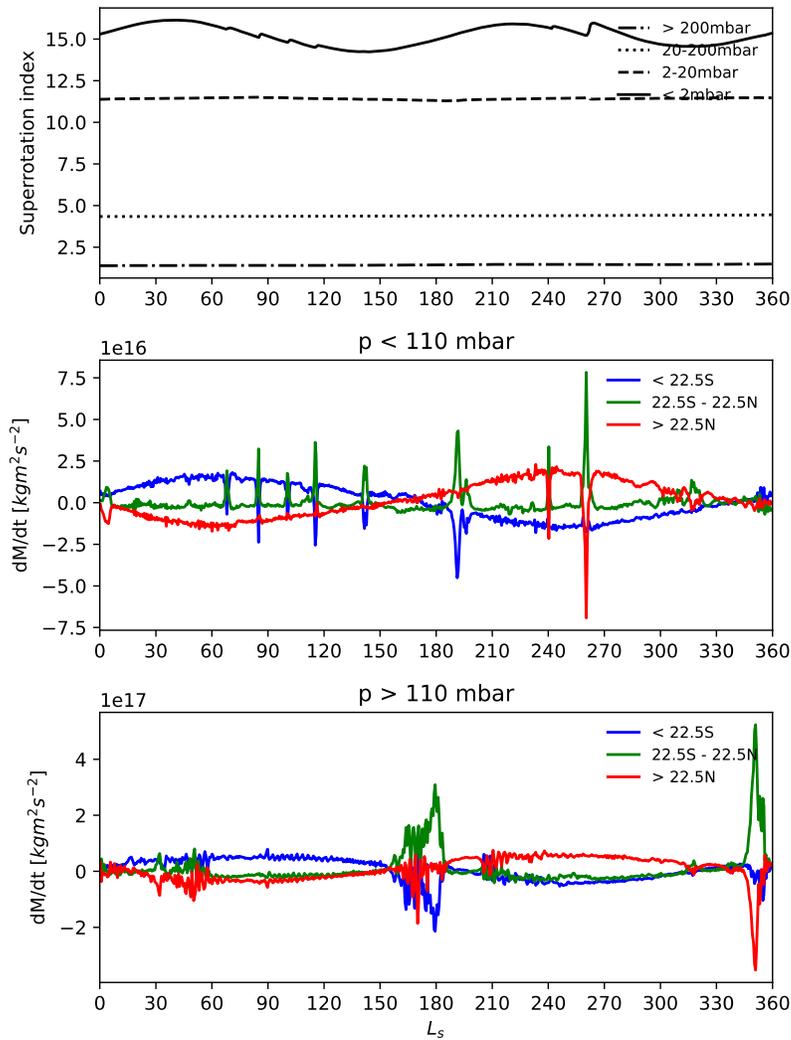

Figure 1. Superrotation indices (top), rate change of angular momentum above 110 mb (middle) and rate change of angular momentum below 110mb (bottom) during one Titan year. Here $p = 110\ mb$ is the pressure near the tropopause. The blue, green and red lines correspond to the southern, equatorial and northern regions respectively. The simulation starts from the spun-up state (year 75) of Newman et al. [2011].

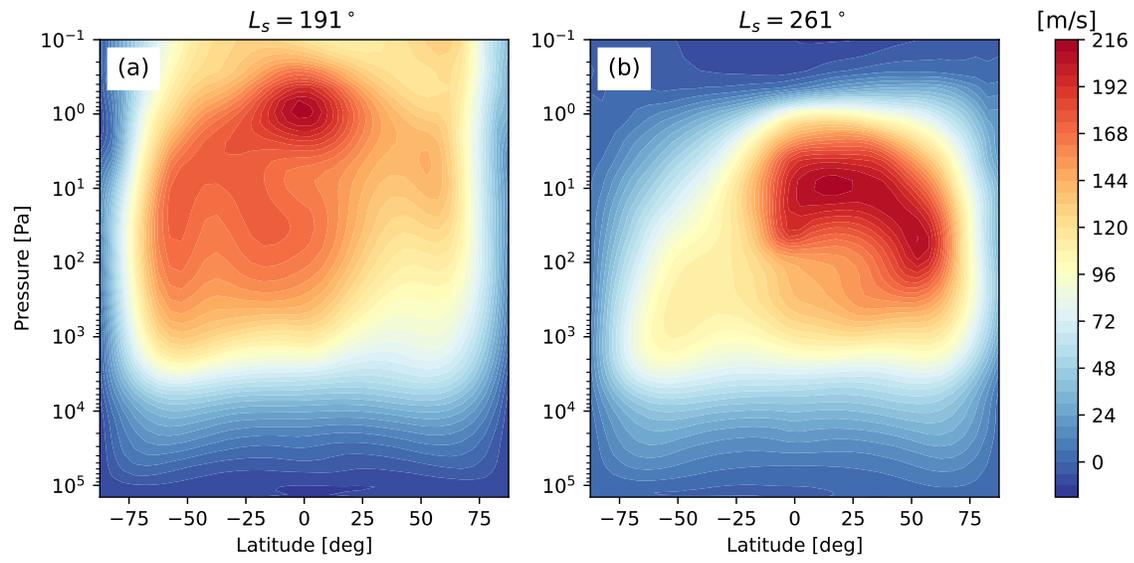

Figure 2. Zonal mean zonal wind near (a) $L_s = 191°$ and (b) $L_s = 261°$.

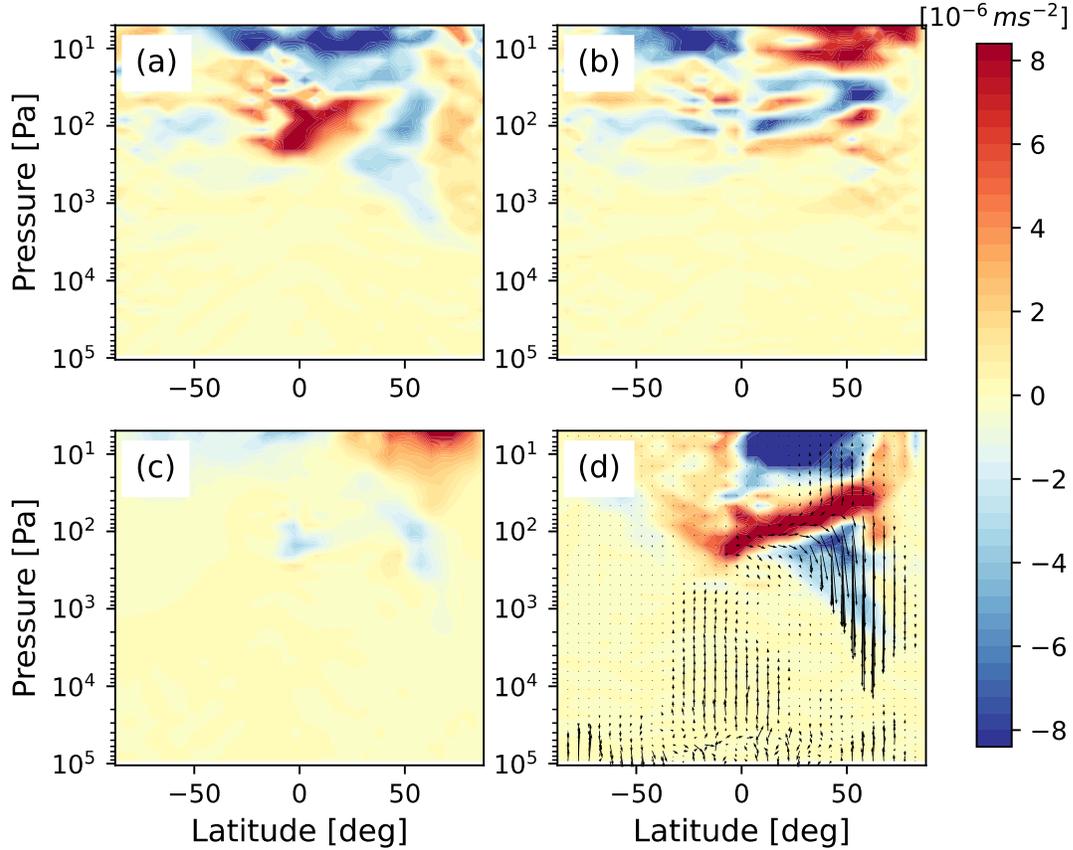

Figure 3. Time-averaged jet acceleration terms from the transformed Eulerian mean diagnostics near $L_s = 261°$. Panels (a), (b), (c) and (d) correspond to the term 1 (total acceleration $\frac{\partial \overline{u}}{\partial t}$), term 2 (Coriolis effect on the meridional residual mean wind), term 3 (vertical advection of zonal momentum due to residual mean circulation), and term 4 (divergence of Eliassen-Palm flux) respectively. The arrows in (d) are the horizontal component ($F_\phi$) and vertical component ($F_z$) of the Eliassen-Palm flux, with the latter scaled by $\pi a/\Delta z$, where $a$ is the radius of Titan and $\Delta z$ is the layer thickness.

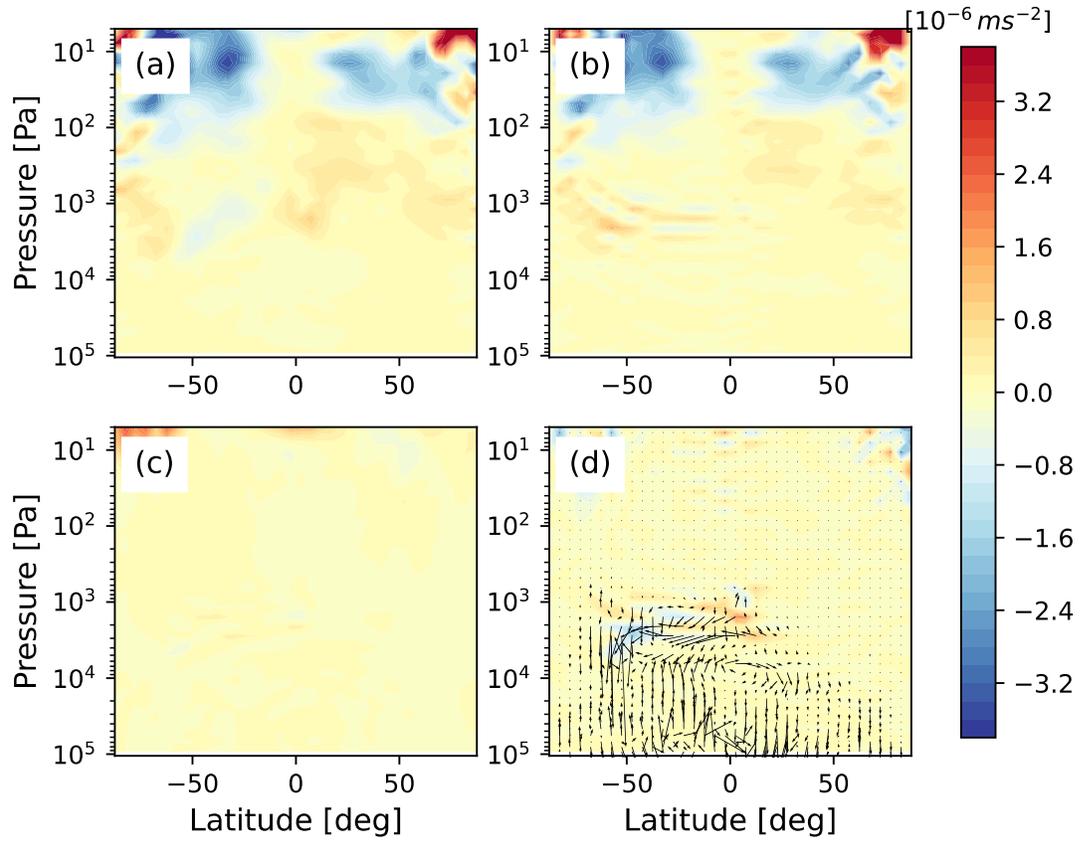

Figure 4. Similar to Figure 3 except it is for TEM diagnostics near $L_s = 191°$.

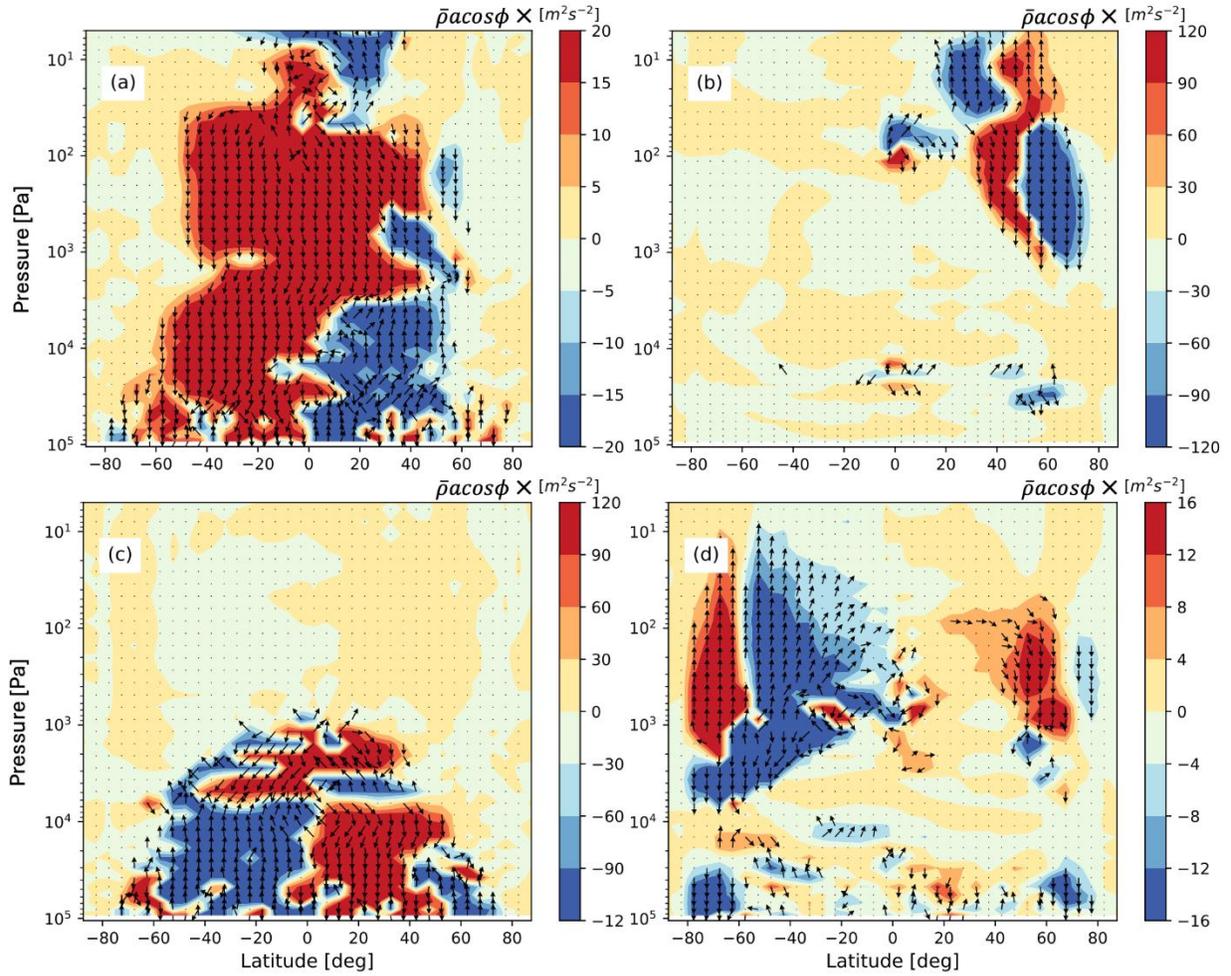

Figure 5. Vertical EMF $\overline{u'w'}$ (colored contours) and EP flux vectors (arrows) for $L_s = 261°$ transfer event (a, b) and $L_s = 191°$ transfer event (c, d). The red (blue) color means upward (downward) flux of eastward eddy momentum. The EMF $\overline{u'w'}$ shown here is multiplied by $\bar{\rho}a\cos\phi$. Only zonal wavenumber 1, the dominant wave mode responsible for zonal wind acceleration/deceleration, is shown for better illustration. Panels (a) and (c) are for prograde waves, and (b) and (d) are for retrograde waves. The EP flux vectors are normalized by their magnitudes over the regions with strong vertical EMF for better visibility of orientations.

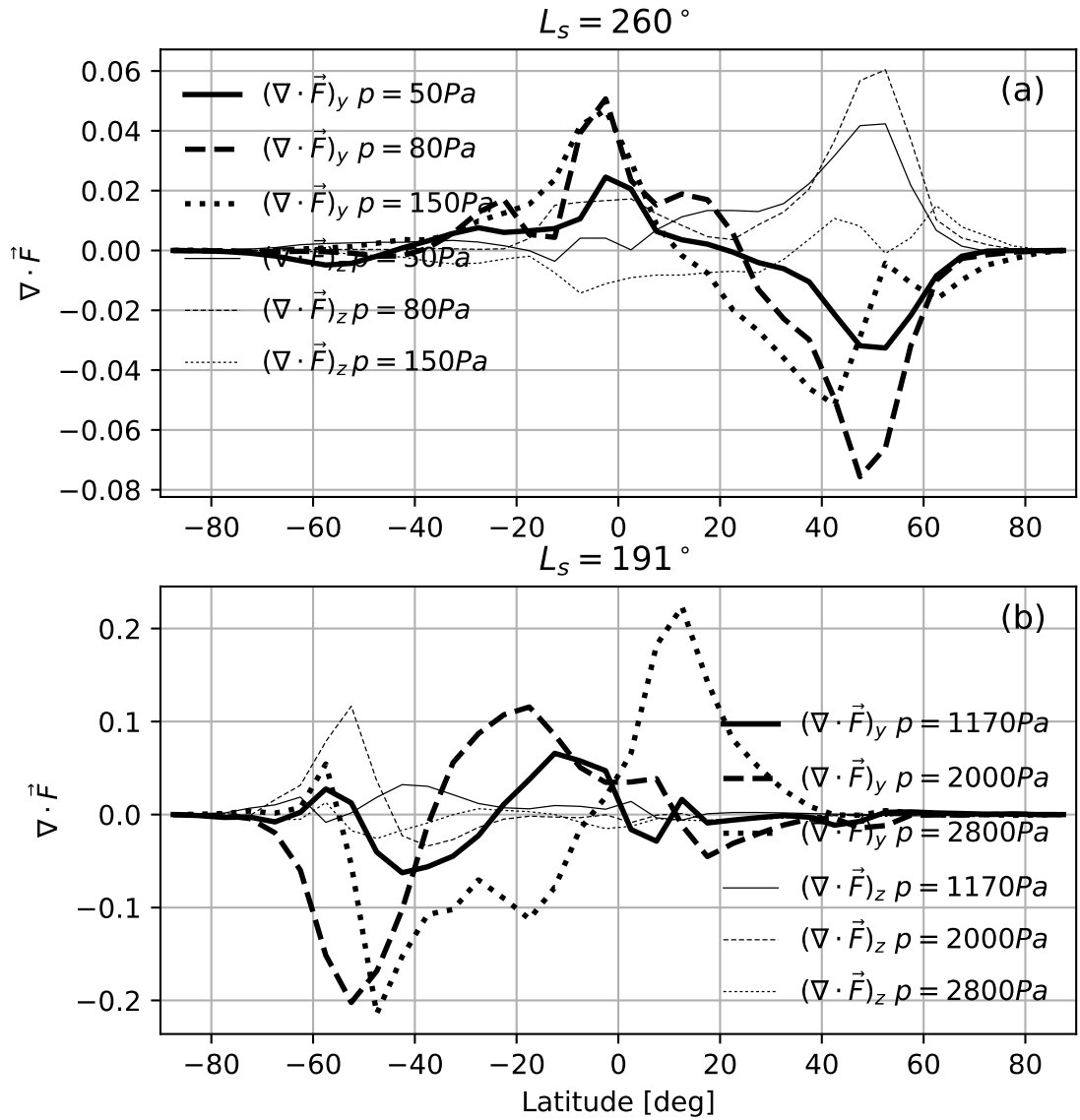

Figure 6. Divergence of the horizontal (thick black lines) and vertical (thin black lines) components of EP fluxes at selected pressure levels for $L_s = 261°$ transfer event (a) and $L_s = 191°$ transfer event (b).

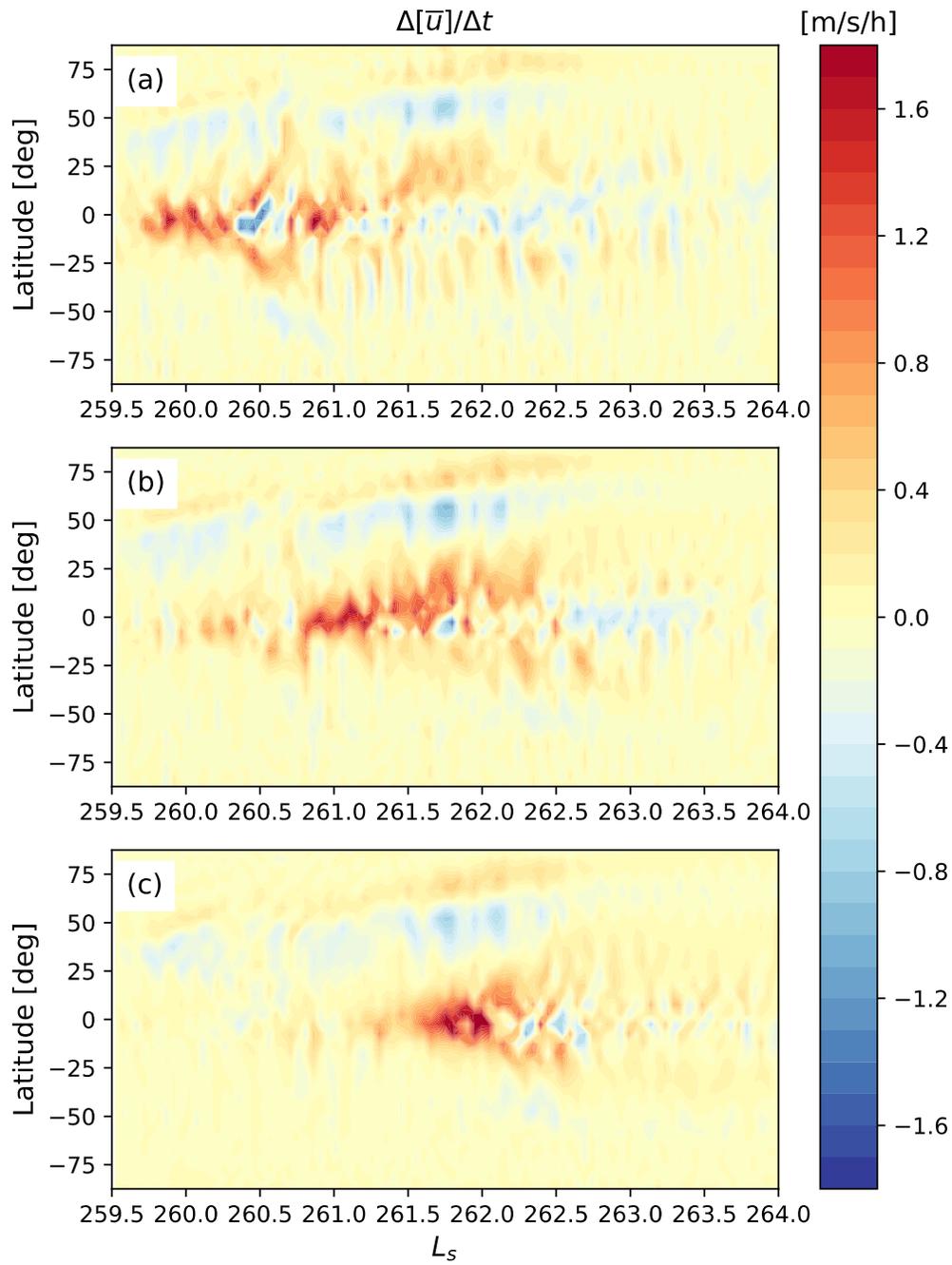

Figure 7. Total acceleration of zonal wind per Titan hour at (a) 50 Pa, (b) 80 Pa and (c) 150 Pa during the $L_s = 261°$ transfer event.

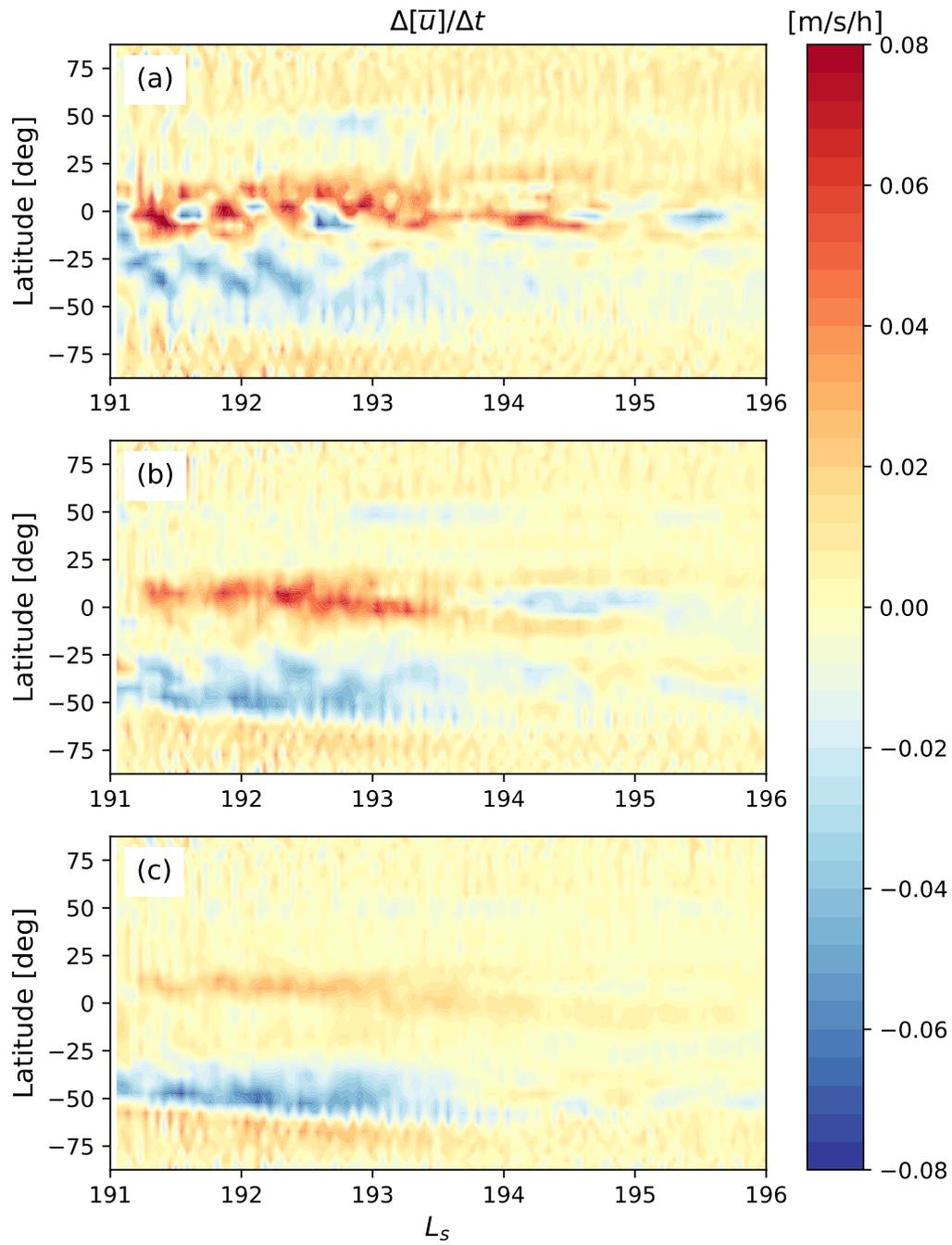

Figure 8. Total acceleration of zonal wind per Titan hour at (a) 1170 Pa, (b) 2000 Pa and (c) 2800 Pa during the $L_s = 191°$ transfer event.

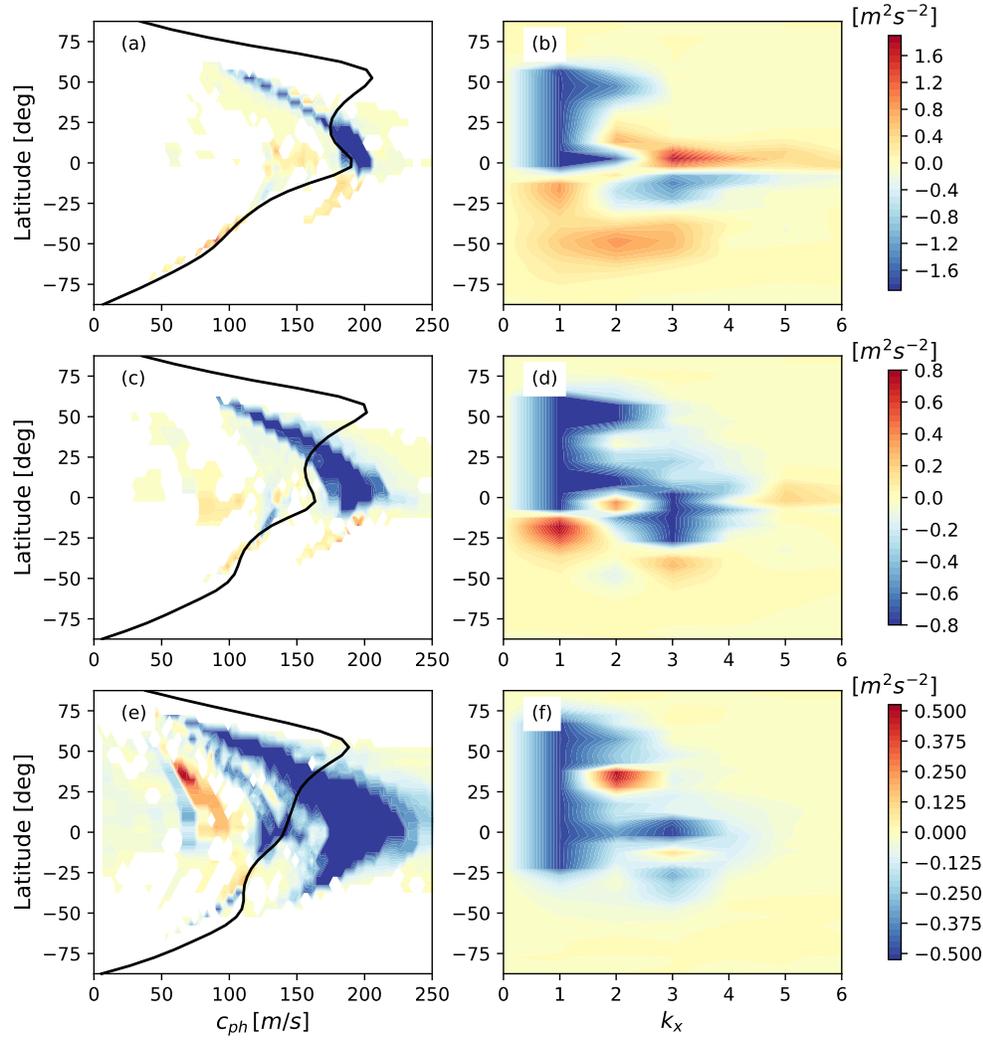

Figure 9. Cospectral analysis of horizontal EMF $u'v'$ near the $L_s = 261.5°$. Panels (a), (c) and (e) show the PSD of EMF as a function of phase speed and latitude (summed over all zonal wavenumbers) at $p$ = 50, 80 and 150 Pa respectively; panels (b), (d) and (f) show the PSD of EMF as a function of zonal wavenumber and latitude (summed over all frequencies) at the corresponding pressure levels. The blue color means southward flux of eastward momentum, and red color means northward flux of eastward momentum. Note that the absolute values of EMF are limited to a smaller range, so the mid-latitude features are more visible.

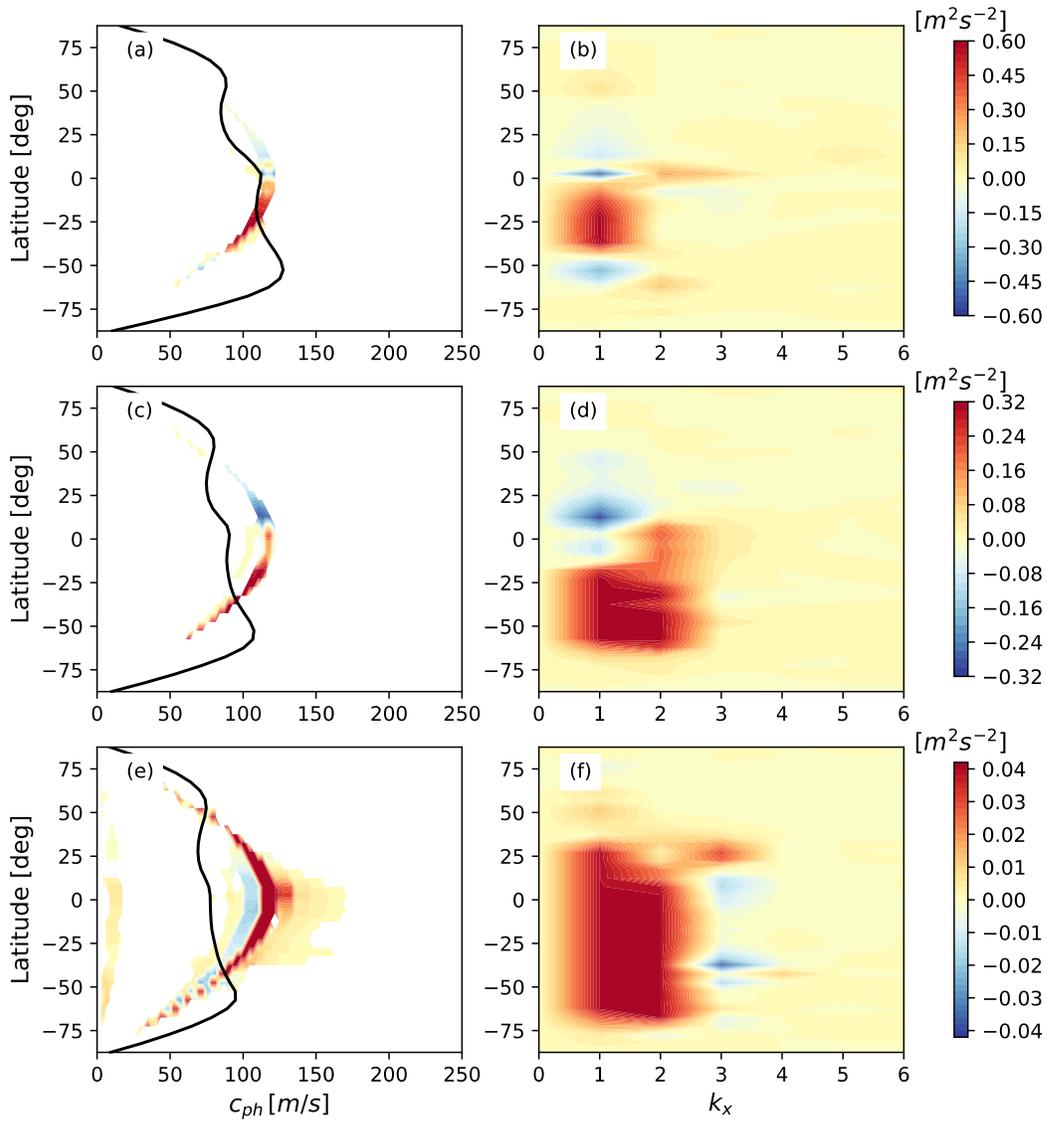

Figure 10. Similar to Figure 9 except they are for the $L_s = 191°$ transfer event. Panels (a)/(b), (c)/(d) and (e)/(f) are at $p$ = 1170, 2000 and 2800 Pa respectively.

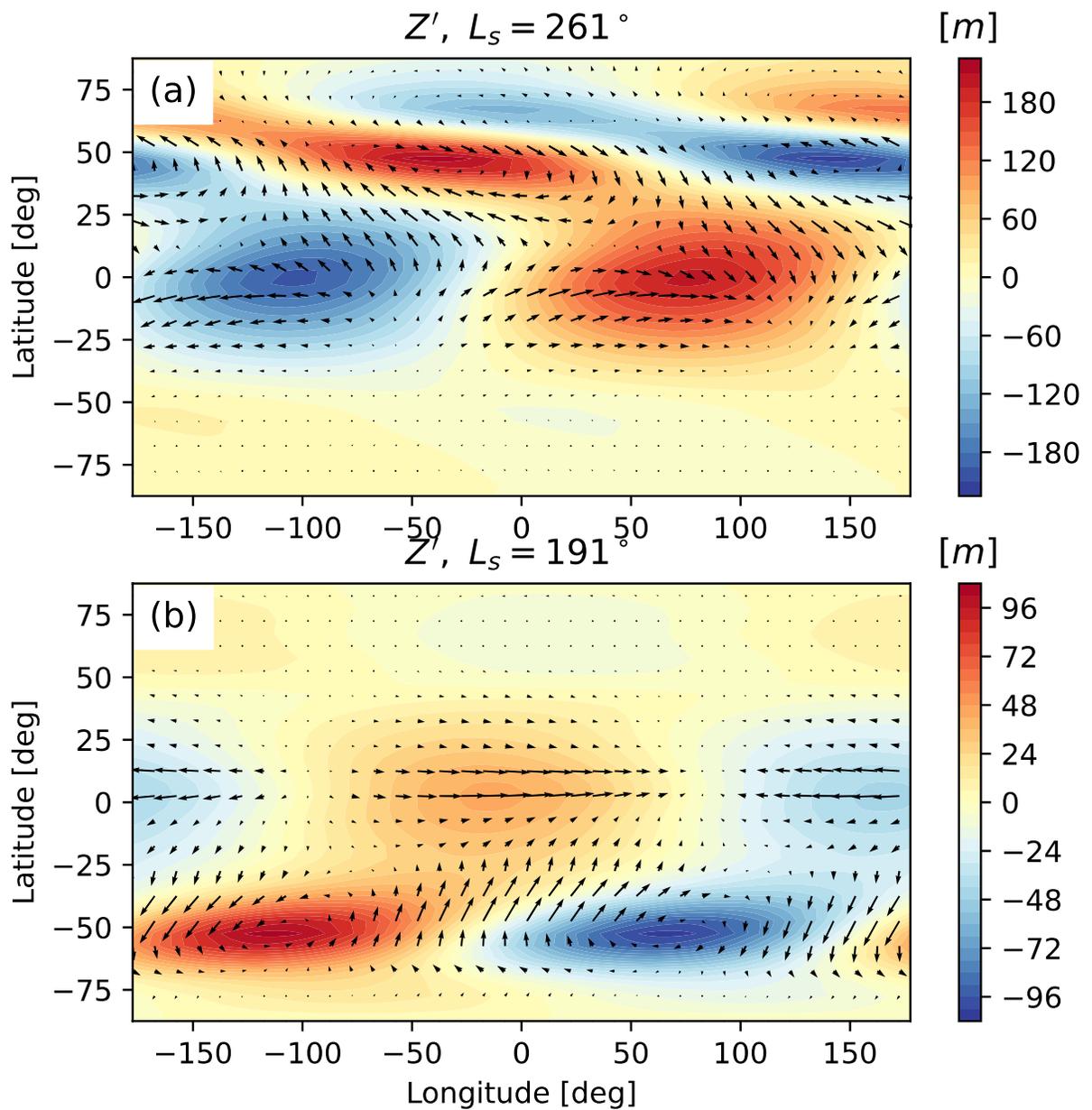

Figure 11. Geopotential anomaly Z′ (color contours) and eddy wind vectors $(u', v')$ (arrows) for zonal wavenumber 1 component at p=80 Pa for $L_s = 261°$ (a) and p=2000 Pa for $L_s = 191°$ (b).

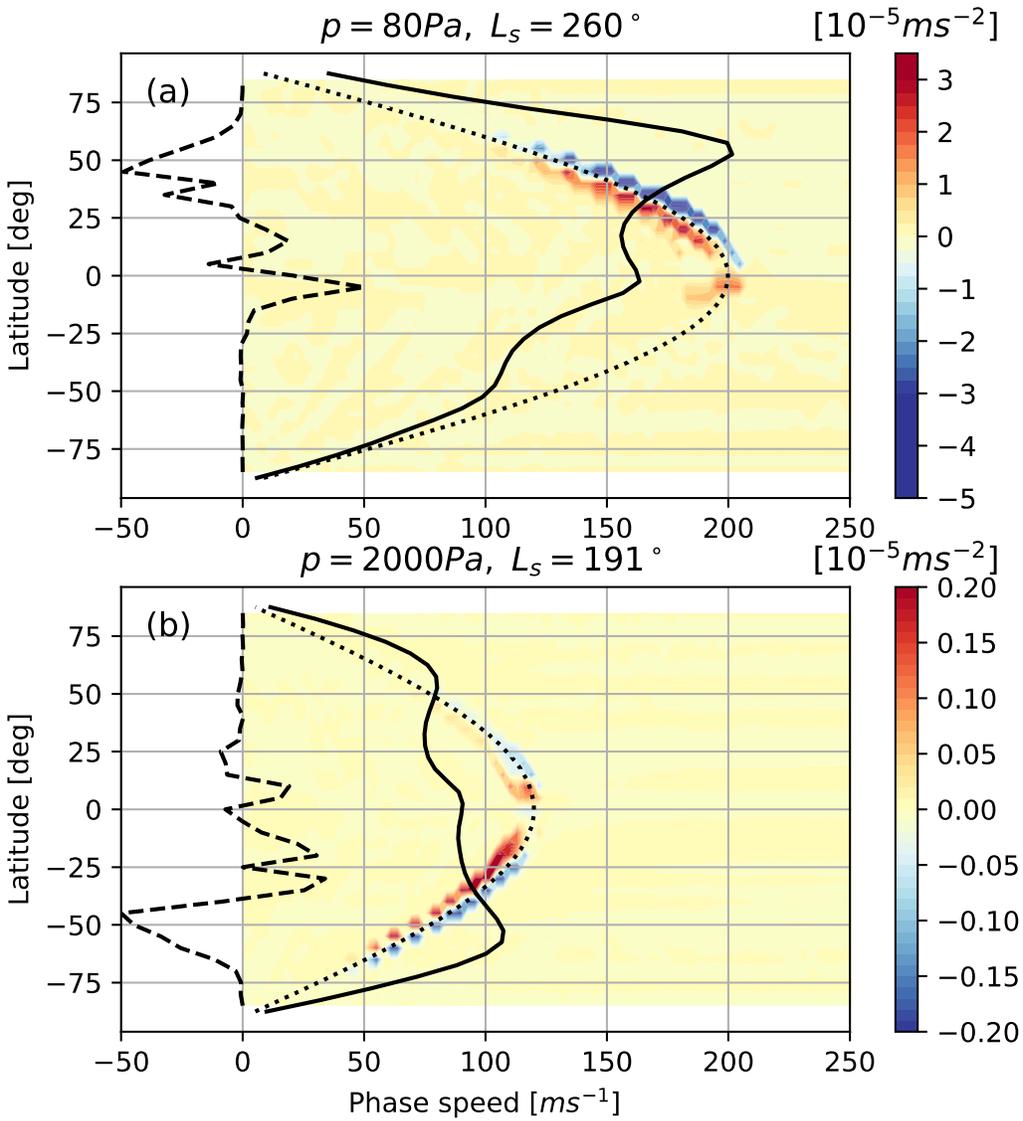

Figure 12. Convergence of horizontal eddy momentum (colored contours) and total zonal acceleration summed over all phase speeds (the dashed lines) near $L_s = 260°$ (a) and $L_s = 191°$ (b). The total zonal acceleration is scaled for better visibility. The solid lines are zonal mean zonal wind as references. The dotted lines are the constant angular velocity $c_0 cos(\phi)$, where $c_0$ is the typical phase speed at equator and $\phi$ is the latitude. The east-westward acceleration correlates well with the intrinsic speed of the dominant wave mode (zonal wavenumber 1).

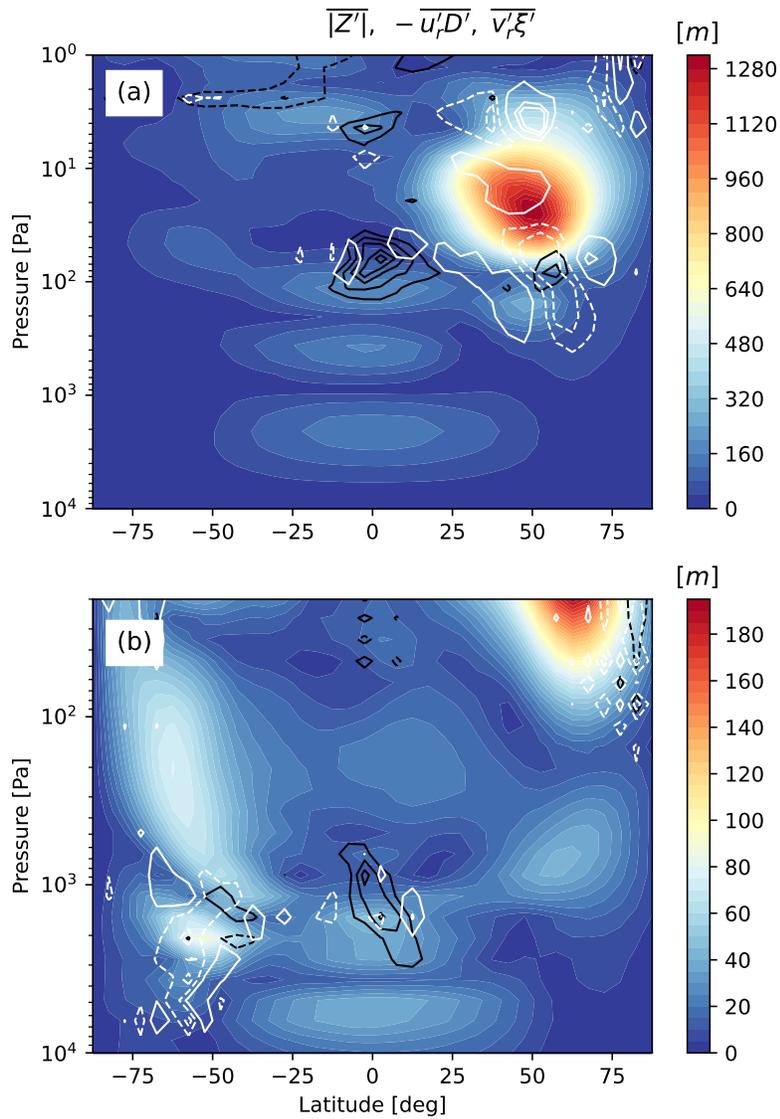

Figure 13. Divergent and rotational contributions to the EMF convergence at zonal wavenumber 1 near $L_s = 260°$ (a) and $L_s = 191°$ (b). The colored contours are the zonal mean of the absolute geopotential anomaly $\overline{|Z'|}$ (root mean square value) with zonal wavenumber 1. The black and white contour lines are the divergent component $-\overline{u_r' D'}$ and rotational component $\overline{v_r' \zeta'}$ respectively. The solid (dashed) lines mean positive (negative) values.

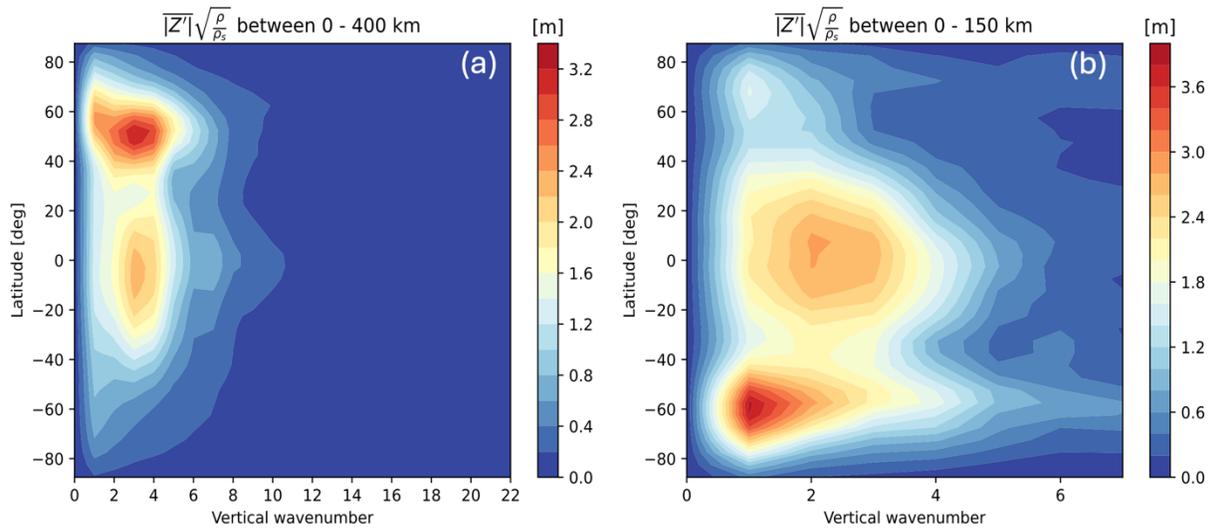

Figure 14. Time-averaged amplitude of density-scaled geopotential anomaly $\overline{|Z'|}$ as a function of vertical wavenumber and latitude for (a) $L_s = 261°$ and (b) $L_s = 191°$ transfer events. The geopotential anomaly is scaled by $\sqrt{\rho/\rho_s}$, and the amplitude is averaged over one Titan day, $\rho$ and $\rho_s$ are the air density and the air density at the ground level respectively.

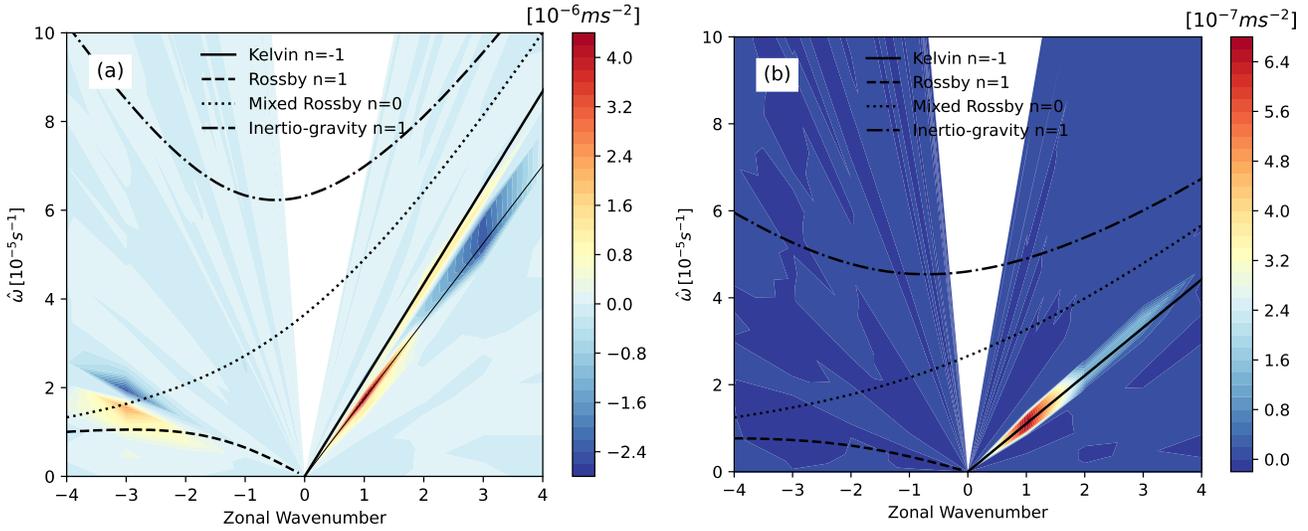

Figure 15. PSD of equatorial EMF convergence (color contours) as a function of zonal wavenumber and intrinsic wave frequency at $\phi = 7.5°$ S and P = 80 Pa for $L_s = 261°$ transfer event (a) and at $\phi = 7.5°$ N and P = 2000 Pa for $L_s = 191°$ transfer event (b). The red color means $-\frac{\partial u'v'}{\partial y} > 0$ and the blue color means $-\frac{\partial u'v'}{\partial y} < 0$ respectively. The solid, dash and dotted lines correspond to the Kelvin wave (n=-1), Rossby wave (only mode n=1 is shown, larger n yields smaller frequency), mixed Rossby-gravity wave (n=0) and inertia-gravity waves (only mode n=1 is shown, larger n yields larger frequency), where n is the meridional mode number. These lines are calculated using the dominant vertical wavenumber and wavelength ($\lambda_z \sim 100\ km$ for $L_s = 261°$ and $\lambda_z \sim 50\ km$ for $L_s = 191°$) obtained from 1D FFT of temperature field at equator (see Fig. 14). The thin black line in (a) is the Kelvin wave calculated with $\lambda_z \sim 80\ km$. The wide space in the figures corresponds to the out-of-range intrinsic phase speeds.

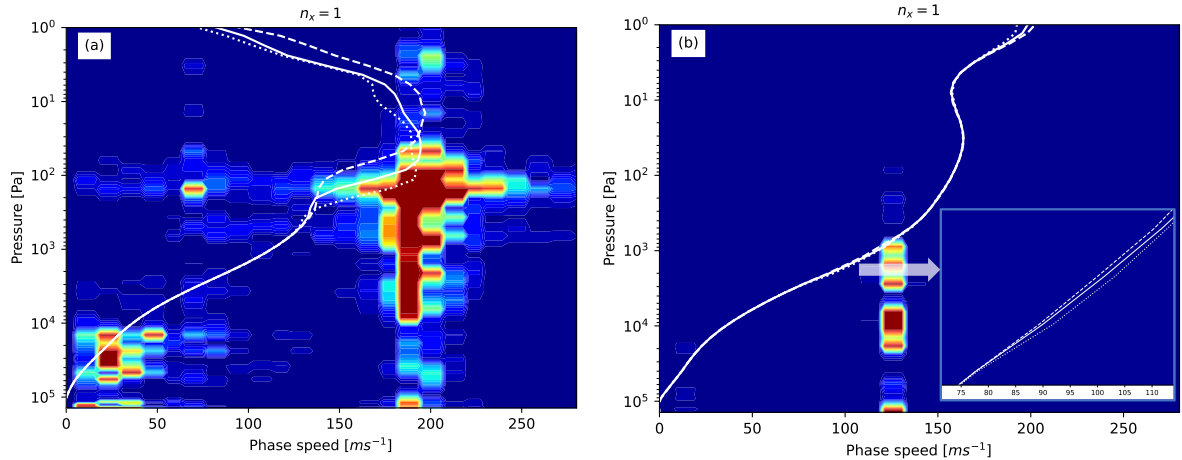

Figure 16. PSD of EMF as a function of phase speed and pressure for (a) $L_s = 261.5°$ and (b) $L_s = 192.3°$ at the equator. The PSD is multiplied by $\rho/\rho_0$ to factor in the growth of wave amplitude due to decreasing density with altitude. Also, it is taken as absolute value for better visibility. The green-red color represents non-zero PSD and blue color represents zero PSD. Only zonal wavenumber 1 is shown. The white lines are zonal mean zonal wind at equator. The dashed, solid and dotted lines in (a) correspond to $L_s = 260.3°$, $L_s = 261.5°$ and $L_s = 262.1°$ respectively. The dashed, solid and dotted lines in (b) corresponds to $L_s = 191.2°$, $L_s = 192.3°$ and $L_s = 193.4°$ respectively. The lines in (b) shows zonal wind is increased by a few meters per second during these time period between 1000 Pa and 3000 Pa (see the zoomed-in region in the lower right corner).

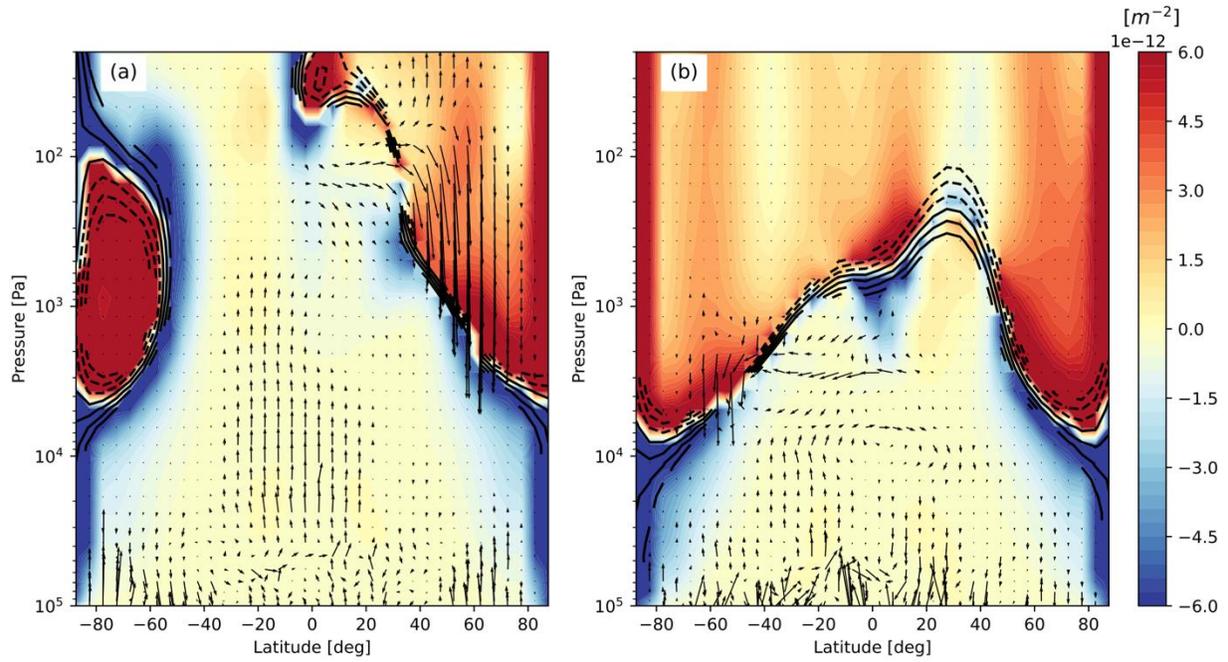

Figure 17. Refractive index $\frac{f^2}{N^2} n_{ref}^2$ for (a) $L_s = 261.5°$ and (b) $L_s = 192.3°$. The black contour lines show the regions near critical lines where the intrinsic phase speed $c - \bar{u}$ is within the range between $\pm 10 ms^{-1}$ (dashed lines for negative values and solid lines for positive values). The arrows are same EP flux vectors as those in Fig. 3d and Fig. 4d.

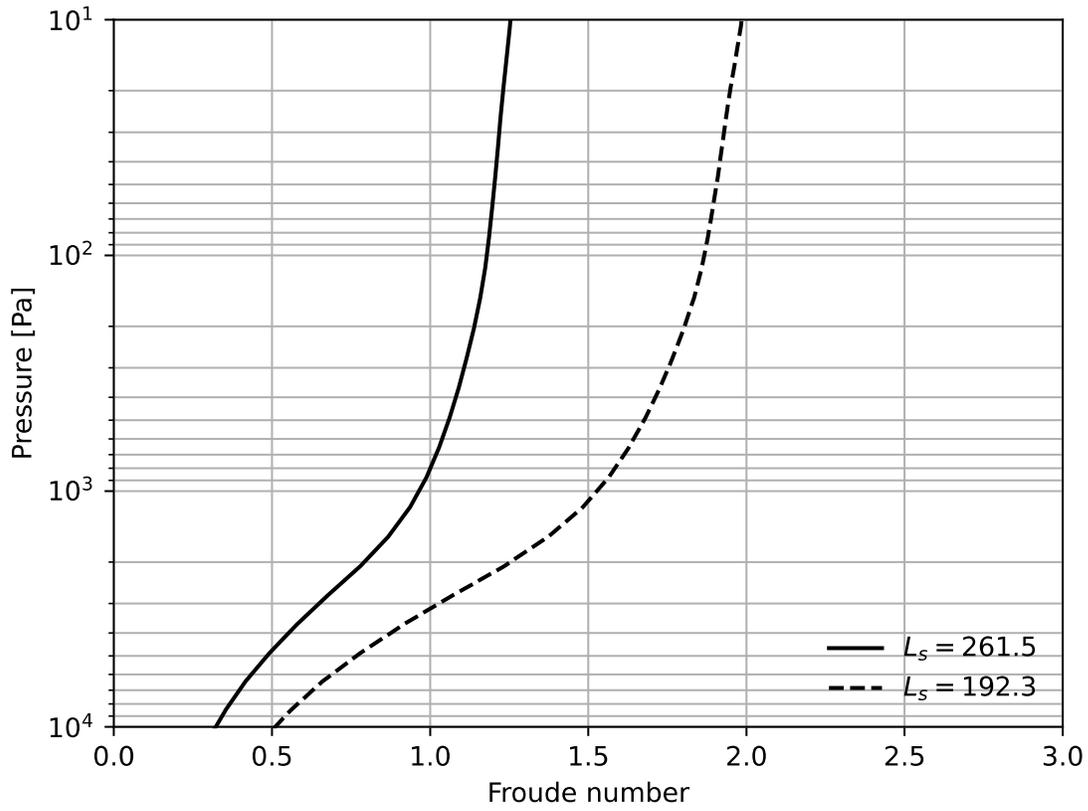

Figure 18. Froude number as a function of pressure for $L_s = 261.5°$ (solid line) and $L_s = 192.3°$ (dashed line).

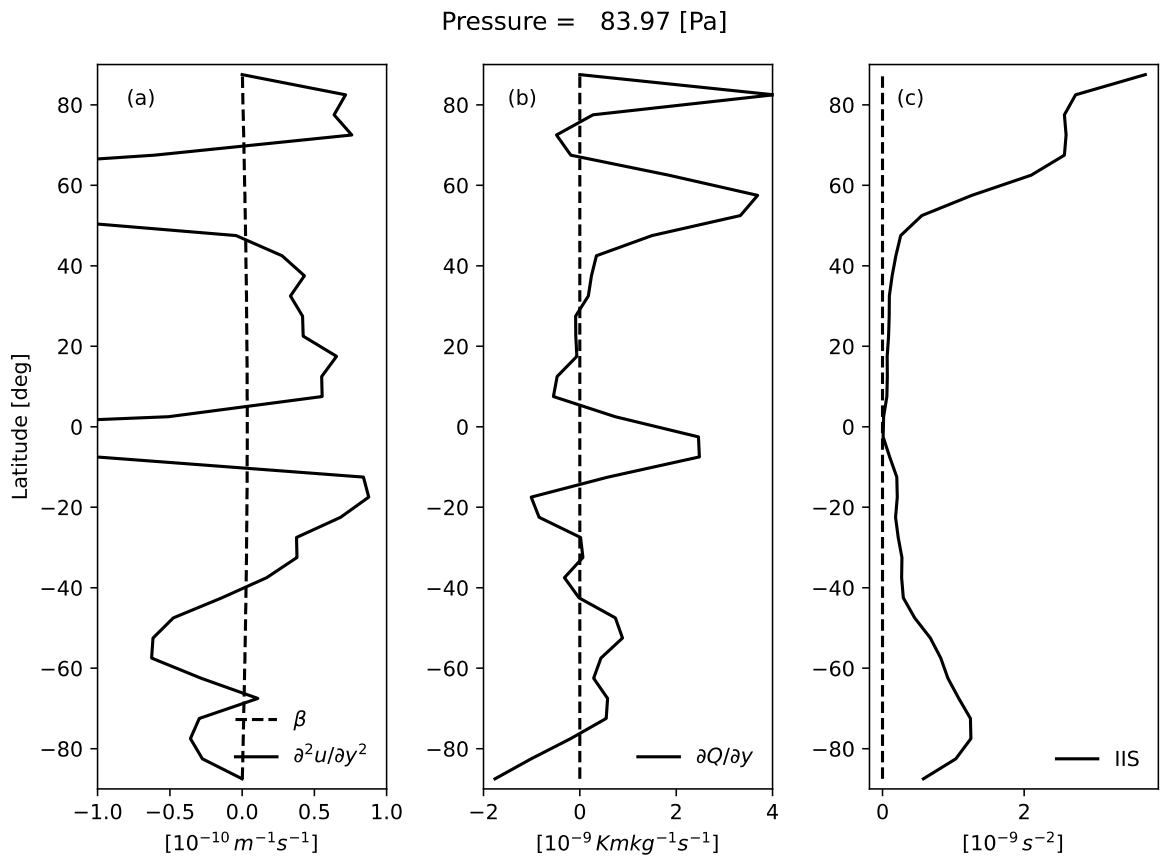

Figure 19. Quantities associated with (a) 2D barotropic, (b) 3D barotropic, and (c) inertial instability criteria, near p=80 Pa at $L_s = 261.5°$. Note that β in (a) is an order of magnitude smaller than the absolute value of wind curvature $\partial^2 u/\partial y^2$, which makes β appear to be zero.

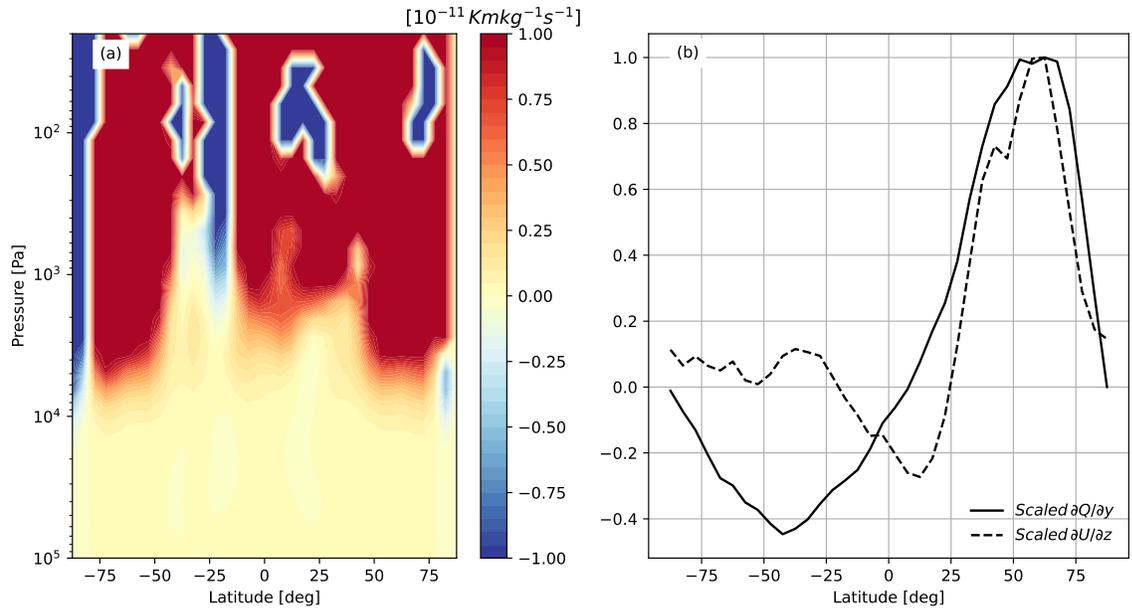

Figure 20. Zonal mean meridional IPV gradient (a) and the meridional QGPV gradient (solid lines) and vertical zonal wind shear (dashed lines) at the ground level near $L_s = 261.5°$ (b). Both quantities in (b) are scaled by their maximum values for better illustration.

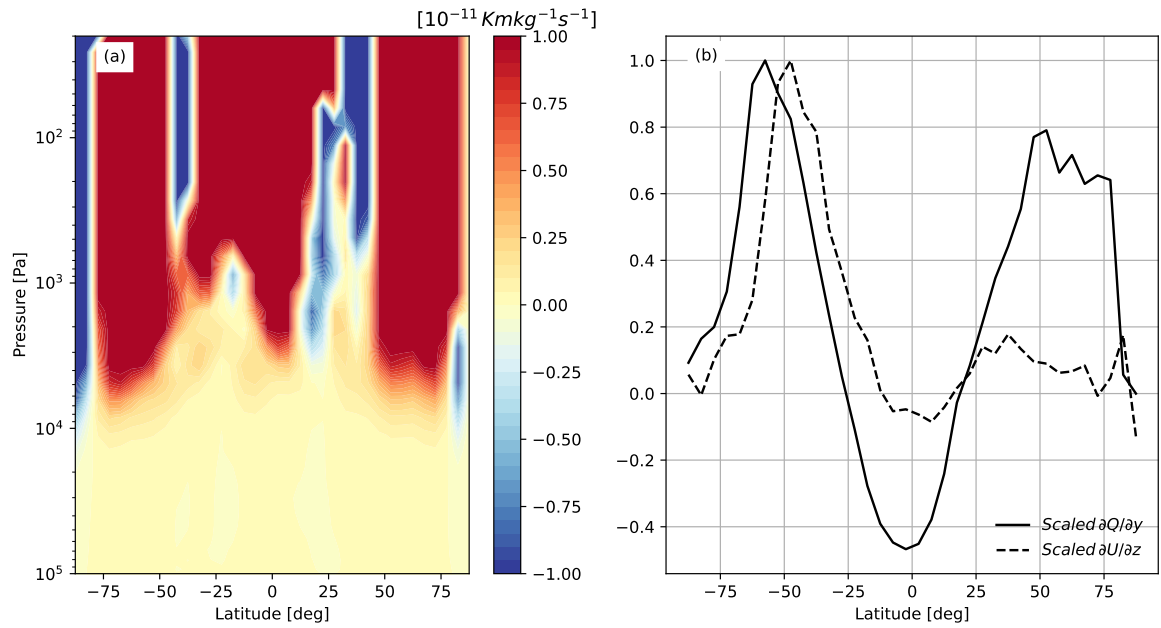

Figure 21. Same as Fig. 20 but for $L_s = 192.3°$